\documentclass[prd,aps,showpacs,superscriptaddress,nofootinbib,eqsecnum,amsfonts,amsmath,twocolumn,preprintnumbers,floatfix,floats]{revtex4}

\usepackage{epsfig}
\usepackage{bm}% bold math
\usepackage{graphics}
\usepackage{amssymb}
\usepackage[usenames]{color}
\usepackage{hyperref}

\def\bea{\begin{eqnarray}}
\def\eea{\end{eqnarray}}
\def\vt{\vartheta}

\begin{document}

%%%%%%%%%%%%%%%%%% For bibtex %%%%%%%%%%%%%%%%
\newcommand{\araa}{{Ann.\ Rev.\ Astron.\ Astrophys.}}
\newcommand{\apjl}{{Astrophys.\ J.\ Lett.}}
\newcommand{\mnras}{{Mon.\ Not.\ Roy.\ Astron.\ Soc.}}
\newcommand{\aap}{{Astron. Astrophys.}}
%%%%%%%%%%%%%%%%%%%%%%%%% define some new commands %%%%%%%%%%%%%%%%%%%%%%%%%%%
\newcommand{\rhat}{\hat{r}}
\newcommand{\iotahat}{\hat{\iota}}
\newcommand{\phihat}{\hat{\phi}}
\newcommand{\h}{\mathfrak{h}}
\newcommand{\be}{\begin{equation}}
\newcommand{\ee}{\end{equation}}
\newcommand{\ber}{\begin{eqnarray}}
\newcommand{\eer}{\end{eqnarray}}
\newcommand{\fmerg}{f_{\rm merg}}
\newcommand{\fcut}{f_{\rm cut}}
\newcommand{\fring}{f_{\rm ring}}
\newcommand{\cA}{\mathcal{A}}
\newcommand{\ie}{i.e.}
\newcommand{\df}{{\mathrm{d}f}}
\newcommand{\rmi}{\mathrm{i}}
\newcommand{\rmd}{\mathrm{d}}
\newcommand{\rme}{\mathrm{e}}
\newcommand{\dt}{{\mathrm{d}t}}
\newcommand{\pj}{\partial_j}
\newcommand{\pk}{\partial_k}
\newcommand{\psifl}{\Psi(f; {\bm \lambda})}
\newcommand{\hp}{h_+(t)}
\newcommand{\hc}{h_\times(t)}
\newcommand{\Fp}{F_+}
\newcommand{\Fc}{F_\times}
\newcommand{\Ylm}{Y_{\ell m}^{-2}}
\def\no{\nonumber \\ & \quad}
\def\noQ{\nonumber \\}
\newcommand{\mc}{M_c}
\newcommand{\vek}[1]{\boldsymbol{#1}}
\newcommand{\vdag}{(v)^\dagger}
\newcommand{\bvtheta}{{\bm \vartheta}}
\newcommand{\btheta}{{\bm \theta}}
\newcommand{\brho}{{\bm \rho}}
\newcommand{\pa}{\partial_a}
\newcommand{\pb}{\partial_b}
\newcommand{\Psieff}{\Psi_{\rm eff}}
\newcommand{\Aeff}{A_{\rm eff}}
\newcommand{\deff}{d_{\rm eff}}
\newcommand{\corr}{\mathcal{C}}
\newcommand{\bvthat}{\hat{\mbox{\boldmath $\vt$}}}
\newcommand{\bvt}{\mbox{\boldmath $\vt$}}
\newcommand{\brhohat}{\hat{\mbox{\boldmath $\rho$}}}
\newcommand{\bsigmahat}{\hat{\mbox{\boldmath $\sigma$}}}
\newcommand{\bsigma}{\mbox{\boldmath $\sigma$}}
\newcommand{\comment}[1]{{\textsf{#1}}}
\newcommand{\dipongkar}[1]{\textcolor{magenta}{\textit{Dipongkar: #1}}}
\newcommand{\sukanta}[1]{\textcolor{blue}{\textit{#1}}}
\newcommand{\UO}{Department of Physics,
1274 University of Oregon, Eugene, OR 97403-1274, U.S.A\\}
\newcommand{\WSU}{Department of Physics \& Astronomy, Washington State University,
1245 Webster, Pullman, WA 99164-2814, U.S.A. \\}
\newcommand{\UWM}{Department of Physics, University of Wisconsin-Milwaukee, 1900 E. Kenwood Blvd., Milwaukee, WI 53211, U.S.A.\\}
\newcommand{\MSU}{Department of Physics, Montana State University, Bozeman, MT 59717-3840, U.S.A.\\
{\textcolor{red} {(DRAFT: Dated \today.)}}
}
\newcommand{\LIGOCaltech}{LIGO Laboratory, California Institute of Technology, 
Pasadena, CA 91125, U.S.A.}
\newcommand{\TAPIR}{Theoretical Astrophysics, California Institute of Technology, 
Pasadena, CA 91125, U.S.A.}
%%%%%%%%%%%%%%%%%%%%%%%%%%%%%%%%%%%%%%%%%%%%%%%%%%%%%%%%%%%%%%%%%%%%%%%%%%%%%%

\title{Improved Coincident and Coherent Detection Statistics for Searches for Gravitational Wave Ringdown Signals}

%\preprint{LIGO-T1300343}  [old]
% https://dcc.ligo.org/LIGO-T1300343
\preprint{LIGO-P1300083}

\author{Dipongkar Talukder}
\email{talukder@uoregon.edu}
\affiliation{\UO}

\author{Sukanta Bose}
\email{sukanta@wsu.edu}
\affiliation{\WSU}

\author{Sarah Caudill}
\email{scaudill@gravity.phys.uwm.edu}
\affiliation{\UWM}

\author{Paul T. Baker}
\email{pbaker@physics.montana.edu}
\affiliation{\MSU}


\begin{abstract}

We study an improved method for detecting gravitational wave (GW) signals from perturbed black holes by earth-based detectors in the quest for searching for intermediate-mass black holes (IMBHs). Such signals, called ringdowns, are damped sinusoids whose frequency and damping constant can be used to measure a black hole's mass and spin. Utilizing the output from a matched filter analysis pipeline, we present an improved statistic for the detection of a ringdown signal that is found to be coincident in multiple detectors. The statistic addresses the non-Gaussianity of the data without the use of an additional signal-based waveform consistency test. We also develop coherent network statistics to check for consistency of signal amplitudes and phases in the different detectors with their different orientations and signal arrival times. We find that the detection efficiency can be improved at least by a few tens of percent by applying these multi-detector statistics primarily because of the ineffectiveness of single-detector based discriminators of non-stationary noise, such as the chi-square test, in the case of ringdown signals studied here.

\end{abstract}
%%%%%%%%%%%%%%%%%%%%%%%%%%%%%%%%%%%%%%%%%%%%%%%%%%%%%%%%%%%%%%%%%%%%%%%%%%%%%%
\maketitle

\section{Introduction}
\label{sec:intro}

Several black holes (BHs) are known to exist with masses as small as about three suns, e.g., IGR J17091-3624, to as large as that of the supermassive black hole in M87, with a mass of about $6.4\times 10^9\, \mbox{M}_\odot$~\cite{0004-637X-489-2-579,0004-637X-700-2-1690}. But until recently, no BH was known to exist with a mass between that of a stellar-mass BH, with mass up to several tens of times the mass of sun, and a super-massive BH, with mass 
%$> 10^5 \,{\rm M}_\odot$. 
less than $10^5 \,{\rm M}_\odot$. 
This wide chasm in the BH mass range is predicted to be populated by IMBHs and has been the subject of debate due to the lack of evidence for their existence. 
% That situation changed fundamentally with the detection of a variable X-ray source 
That situation changed with the detection of a variable X-ray source, HLX-1, with a maximum 0.2-10 keV luminosity and lower mass limit of $\sim 500\,{\rm M}_\odot$ in the spiral galaxy ESO 243-49~\cite{Farrell:2010bf}. If a population of binary IMBHs with total mass less than several hundred solar masses exists and merges on a short timescale, GWs will be emitted in the band of the earth-based detectors like LIGO and Virgo. It is also possible for a perturbed Kerr black hole, with a mass between a few to thousands of solar masses, to radiate away the perturbation in GWs as it rings down to a stationary state.

The detection of GWs from IMBHs will have important consequences for theories about the formation of supermassive black holes and the dynamics and evolution of globular clusters~\cite{2041-8205-713-1-L41,0004-637X-750-1-31}. 
%The following sentence is updated below to make it clearer that we are talking about IMBHs and not SMBHs: They will also allow the direct measurement of their masses and spin parameters.
It will also allow the direct measurement of the masses and spins of IMBHs. The merger and ringdown phases of the GW signal are important for the detection of IMBH sources because for massive systems the characteristic frequencies of the inspiral phase are outside of the sensitivity band of ground based detectors.

The Laser Interferometer Gravitational-wave Observatory (LIGO) aims to detect gravitational waves from known and unknown sources. Matched filtering is known to be the optimal technique for finding known signals buried in Gaussian and stationary noise~\cite{helstrom1960statistical}. But data from the LIGO detectors exhibits non-Gaussian, non-stationary noise producing many false candidate events in gravitational-wave searches~\cite{0264-9381-27-16-165023}. In order to increase the detection probability, it is therefore necessary to have statistics that are optimal in separating true signals from non-Gaussian noise.  

Here, we study a method of detecting the IMBHs via the {\em ringdown} phase of a GW signal that arises when such a black hole is perturbed. The layout of the paper is as follows. The rest of Sec.~\ref{sec:intro} describes the ringdown signals. In Sec.~\ref{sec:coinc}, we describe the challenges posed by a coincident multi-detector search of such signals. In section~\ref{sec:coherent}, we introduce coherent statistics for the same signals in data from a network involving two or three detectors.

\subsection{The ringdown waveform}\label{ringwaveform}

A BH can be perturbed in a variety of ways, e.g., by the incidence of GWs, by an object falling into it, by the interaction with a companion, by the accretion of matter surrounding it, or by the formation process in a gravitational collapse. There are no normal mode oscillations associated with non-radial perturbations. %This is in contrast with the normal modes of Newtonian gravity, because in GR they are damped by the emission of GWs and, hence, they are called quasi-normal modes. 
Numerical simulations (see for example, Refs.~\cite{PhysRevD.75.124018,PhysRevD.76.064034}) have demonstrated that the fundamental mode, $l=m=2$, dominates the GW emission. Far from the source, the plus and cross polarizations of a ringdown waveform, approximated for the $l=m=2$ mode, can be expressed in terms of the central frequency $f_0\equiv f_{22}$ and the quality factor $Q\equiv~Q_{22}$ for $t>0$ as~\cite{PhysRevD.46.5236,PhysRevD.80.062001,Goggin:2008dz,0264-9381-23-19-S09}
\begin{eqnarray}\label{eq5_4}
h_{+}(t)&=&\frac{{\cal A}}{r}\,(1+\cos^{2}\iota)\,e^{-\frac{\pi f_{0}t}{Q}}\cos(2\pi f_{0}t+\chi)\, ,\\
h_{\times}(t)&=&\frac{{\cal A}}{r}\,(2\cos\iota) ~e^{-\frac{\pi f_{0}t}{Q}}\sin(2\pi f_{0}t+\chi)\, ,
\end{eqnarray}
where ${\cal A}$ is the amplitude of the $l=m=2$ mode, $\chi$ is the initial phase, $\iota$ is the inclination angle and $r$ is the distance to the source. The strain produced in the detector is then
\begin{equation}\label{eq5_5}
h(t)=h_{+}(t-t_{0})F_{+}(\theta,\phi,\psi)+h_{\times}(t-t_{0})F_{\times}(\theta,\phi,\psi)\, ,
\end{equation}
where $t_{0}$ is the signal's arrival time at the detector, $F_{+,\times}$ are the detector antenna-pattern functions~\cite{PhysRevD.63.042003}, $\psi$ is the angle that defines the orientation of the polarization ellipse of the signal 
and $(\theta, \phi)$ are the sky-position angles of the source.

If $\epsilon~(\ll 1)$ is the fraction of a black hole's mass $M$ radiated as gravitational waves, then their strain amplitude is given by
\begin{equation}\label{eq5_7}
{\cal A}=\sqrt{\frac{5}{2}\epsilon}\,\left(\frac{GM}{c^{2}}\right)\,Q^{-1/2}F(Q)^{-1/2}g(\hat{a})^{-1/2}\,,
\end{equation} 
where $\hat{a}$ is a dimensionless spin parameter, $g(\hat{a})=\left[1.5251-1.1568(1-\hat{a})^{0.1292}\right]$ and $F(Q)=1+\frac{7}{24Q^{2}}$~\cite{Goggin:2008dz}. The frequency $f_0$ and quality factor $Q$ of each quasi-normal mode can be related to the black hole mass and spin through a fitting formula~\cite{Leaver09121985,PhysRevD.40.3194,PhysRevD.73.064030,PhysRevD.76.104044,0264-9381-29-9-095016}. For the $l=m=2$ mode, it gives:
\begin{eqnarray}\label{eq5_8}
Q &=& 0.7000+1.4187(1-\hat{a})^{-0.4990}\,,\\
f_{0} &=& \frac{1}{2\pi}\frac{c^{3}}{GM}\left[1.5251-1.1568(1-\hat{a})^{0.1292}\right]\,.
\end{eqnarray}  
Note that while $Q$ is determined by $\hat{a}$ alone, $f_{0}$ is determined by both $M$ and $\hat{a}$. %Figure~\ref{fig:M_fanda_Q} shows how frequency and quality factors are related to mass and spin.

For a given source at a distance $r$, one defines the effective distance as
\begin{equation}\label{eq5_5a}
D_{\mbox{\scriptsize{eff}}}=\frac{r}{\sqrt{F_{+}^{2}(1+\cos^{2}\iota)^{2}/4+F_{\times}^{2}\cos^{2}\iota}}\,,
\end{equation}
%The effective distance is the distance to a GW source as if it were 
which reduces to $r$ for a GW source that is optimally located and oriented. Note that $D_{\mbox{\scriptsize{eff}}} \geq r$.
%For non-optimal location or orientation, $$.
By substituting the expressions of the polarization components given in Eq.~(\ref{eq5_4}) in the GW strain expression (\ref{eq5_5}), one finds that the ringdown waveform of a black hole is approximated by~\cite{PhysRevD.60.022001,PhysRevD.80.062001,0264-9381-23-19-S09},
\begin{equation}\label{eq5_6}
h(t)={\cal A}_{\mbox{\scriptsize{eff}}}\,e^{-\frac{\pi f_{0}(t-t_{0})}{Q}}\cos(2\pi f_{0}(t-t_{0})+\varphi_{0})\, ,
\end{equation}
where $t>t_{0}$,
${\cal A}_{\mbox{\scriptsize{eff}}} \equiv 2{\cal A}/D_{\mbox{\scriptsize{eff}}}$, and
\begin{equation}\label{varphi0}
\varphi_{0} \equiv \chi - \tan^{-1}\left\{\frac{2F_\times\cos\iota}{F_+\left(1+\cos^{2}\iota\right)}\right\}\,.
\end{equation}
We call $\varphi_{0}$ the effective initial phase of the strain signal in a detector. It is determined by the phase $\chi$ in Eq.~(\ref{eq5_4}) as well as a term that depends on the polarization of the signal through $\iota$, and the orientation of the polarization ellipse given by $\psi$. A search can be designed to track the variation of effective initial phases of signals from the same source in various detectors and check for their consistency with the variation in the signal amplitudes in the same detectors. This check is implemented in Sec.~\ref{sec:coherent} in what is termed here as the ``coherent'' search.

\section{A coincident multi-detector search}
\label{sec:coinc}

Several ringdown searches have been carried out in the last few years~\cite{PhysRevD.60.022001,RanaThesis,0264-9381-21-5-047}. In 2009, a 90\%-confidence upper limit was placed on the rate of ringdowns from BHs with masses between $85~\mbox{M}_{\odot}$ and $390~\mbox{M}_{\odot}$ in the local universe, assuming a uniform distribution of sources, of $3.2\times10^{-5}~{\mbox{yr}}^{-1}{\mbox{Mpc}}^{-3}$~\cite{PhysRevD.80.062001,Goggin:2008dz}. This search was carried out on data from the fourth LIGO science run S4, which took place between February 22 and March 24, 2005. We refer to it as the ``S4 ringdown search''. A weakly modeled burst search for GWs from mergers of non-spinning intermediate-mass binary black holes was performed on data from LIGO S5 and Virgo VSR1 science runs. A 90\%-confidence upper limit of $0.13~{\mbox{Myr}}^{-1}{\mbox{Mpc}}^{-3}$ is placed on the rate of non-spinning sources with component masses $m_{1}=m_{2}=88~{\mbox{M}}_{\odot}$~\cite{PhysRevD.85.102004}. 

In the following sections, we review the method used for searching for ringdown GW signals in LIGO and Virgo data from perturbed BHs. We focus on the search for ringdown signals in the data from LIGO S5 and S6 runs and Virgo VSR2 and VSR3 runs~\cite{S5_S6_paper}. We refer to it as the ``S5/S6 ringdown search''. The data analyzed in this search was collected between 4th November 2005 and 20th October 2010 (with commissioning breaks in midway). LIGO comprises two observatory sites: Hanford, WA, (or ``LHO'') hosts two detectors of arm lengths 4 km and 2 km (termed as H1 and H2, respectively) and Livingston, LA, (or ``LLO'') hosts one detector of arm length 4 km (L1)~\cite{LIGO_USA}. The Virgo detector (V1), with 3 km long arms, is located in Cascina, Italy~\cite{Virgo_Europe}. 

\subsection{The coincidence search pipeline}
\label{sec:coincpipe}

\begin{figure}[!h]
\begin{center}
%\vspace{-0.3in}
{\includegraphics[width=0.5\textwidth]{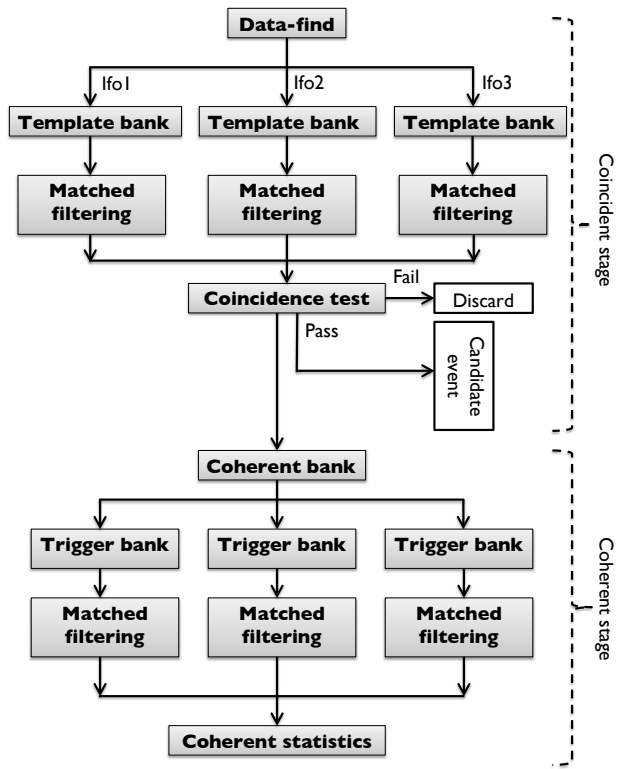}}
%\vspace{-0.6in}
\caption{\label{fig:pipe} A schematic diagram of the coincidence and coherent (see Section~\ref{app:cohpipe}) stages of the ringdown search pipeline. A network of three detectors (named as ifo1, ifo2, and ifo3) are considered as an example of this diagram.}
\end{center}
\end{figure}

The optimal method for finding known signals buried in Gaussian detector noise is to match-filter a detector's output with theoretically modeled waveforms~\cite{helstrom1960statistical}. The ``ringdown search pipeline'', illustrated in Fig.~\ref{fig:pipe}, is a multi-detector data analysis pipeline designed for searching the $l=m=2$ quasi-normal mode of gravitational waves from perturbed black holes~\cite{Goggin:2008dz,PhysRevD.80.062001}. Here we summarize the main steps of its coincident stage. The coherent stage of the pipeline is described in Section~\ref{app:cohpipe}. 

The first stage of the pipeline involves reading in and conditioning the data from each of the detectors. 
%We read in 2176s long segment of the calibrated data. %We read in uncalibrated data at a sampling rate of 16384 Hz, low-pass filter it to remove any power above 4096 Hz as a part of downsampling it to 8192 Hz to reduce the computational cost. The data is then high-pass filtered to remove power below 40 Hz. This is converted to strain by applying the detector response function. The one-sided power spectral density is calculated for each 2176s long segment of the calibrated data. 
The data are then segmented~\footnote{The segmentation of data is discussed in more detail in Ref.~\cite{Brown:2004vh}.} and filtered with a bank of ringdown templates characterized by either mass and spin or frequency and quality factor. Following Refs.~\cite{PhysRevD.80.062001,PhysRevD.60.022001,Goggin:2008dz}, the template used in this search is given by
%\textcolor{red}{Do you want to explain the use of the subscript c in following equation?} 
%\footnote{The metric in Eq.~\eqref{eq5_10} is derived using the damped-sine phase of the waveform, i.e., $h_{s}(t) = e^{-\frac{\pi f_{0}t}{Q}}\sin(2\pi f_{0}t)$.}
\begin{equation}\label{eq5_13}
h_{\scriptsize{\mbox{c}}}(t) = e^{-\frac{\pi f_{0}t}{Q}}\cos(2\pi f_{0}t)\,, \;\;\;\;\;\;0\leq t\leq t_{\text{max}}
\end{equation}
with a length of 10 e-folding times, $t_{\text{max}}=10\tau$, where $\tau=Q/\pi f_{0}$. Here the subscript ${\scriptsize{\mbox{c}}}$ signifies the cosine phase of the waveform. We construct a bank of templates to search over ranges of the two intrinsic parameters of interest~\cite{Owen:1995tm,PhysRevD.80.062001}. Filtering the data $x(t)$ of a single detector with $h_{\scriptsize{\mbox{c}}}$ yields the signal-to-noise ratio (SNR) statistic given by
\begin{equation}\label{eq5_14}
\rho_{\mbox{\scriptsize{c}}}(h_{\mbox{\scriptsize{c}}})= \frac{|\langle x,h_{\mbox{\scriptsize{c}}}\rangle|}{\sqrt{\langle h_{\mbox{\scriptsize{c}}},h_{\mbox{\scriptsize{c}}}\rangle}}=\frac{|Z_{\mbox{\scriptsize{c}}}|}{\sigma_{\mbox{\scriptsize{c}}}} \,,
\end{equation} 
where $Z_{\mbox{\scriptsize{c}}}\equiv\langle x,h_{\mbox{\scriptsize{c}}}\rangle$, $\sigma_{\mbox{\scriptsize{c}}}\equiv\sqrt{\langle h_{\mbox{\scriptsize{c}}},h_{\mbox{\scriptsize{c}}}\rangle}$, and $\langle x,h_{\mbox{\scriptsize{c}}}\rangle$ denotes the noise-weighted inner product of the data and the template:
\be\label{innerprod}
\langle x ,h_{\scriptsize{\mbox{c}}} \rangle = 2 \int_{-\infty}^{\infty}\df\, {\tilde{x}(f)\,\tilde{h}^{*}_{\scriptsize{\mbox{c}}}(f) \over S_{h}(|f|)} \, , 
\ee
where $\tilde{x}(f)$ and $\tilde{h}_{\scriptsize{\mbox{c}}}(f)$ are the Fourier transforms of $x(t)$ and $h_{\scriptsize{\mbox{c}}}(t)$, respectively, and $S_{h}$ is the noise power spectral density (PSD) of the data segment being filtered. The phase of $Z_c$ is sometimes termed as the quadrature phase, which we will return to in Sec.~\ref{sec:coherent}.
%(Note that $\rho_{\mbox{\scriptsize{c}}}$ is devoid of any information about this phase.)

A trigger is generated when the SNR for any template crosses a preset threshold. The threshold is chosen carefully so as to minimize the false dismissal rate for a given false-alarm rate. Once triggers are found in one detector they are checked for parameter consistency and time delay with triggers from other detectors, that were operating concurrently, to increase the confidence level for the presence of a signal. This is commonly known as a coincidence test~\cite{PhysRevD.78.062002}. In the S5/S6 ringdown search, to assess how similar two triggers are in two different detectors and, accordingly, determine if they are coincident, a 3D metric is constructed on the $(f_{0},Q,t)$-space to calculate the distance between their two respective templates~\cite{S5_S6_paper}. This metric is found to be better performing than the 2D metric used in the S4 ringdown search. At this stage we also veto triggers occurring during times when data quality flags are on. These flags mark stretches of data where the astrophysical origin of a trigger is suspect owing to poor data quality, e.g., due to the presence of environmental or instrumental artifacts. Consequently, triggers from these stretches are dropped. On the other hand, triggers that survive this veto are recorded as coincidences and are followed up further because they can be signals.

%CHECK: \begin{figure}[ht!]

\subsection{Tuning the search}

Since the noise in the data stream is nonstationary and non-Gaussian, matched filtering alone 
%is not enough to establish the significance of a trigger as a GW signal. 
does not provide for the best discriminator of a GW signal.
Artifacts in detector noise can often mimic the signals we are searching for, and so a large effort goes into characterizing the noise to best separate those artifacts from potential gravitational wave signals. We estimate the background due to accidental coincidences of noise by time-shifting triggers in one detector relative to those in another by durations greater than the light-travel-time between those detectors. These acausal time-shifts 
%CHECK: minimize the chance 
preclude the possibility of including a truly coincident GW trigger in the background sample. We refer to these as background triggers or slide triggers, as opposed to the in-time coincident triggers (i.e., zero-lag triggers) obtained without such time shifts. Background triggers corresponding to times that are flagged by data quality studies are discarded. 

In order to test the sensitivity of our search to GWs from BHs, we add a large set of simulated signals to the data stream, in software. These are known as ``software injections''. We then run the search pipeline to find them in the noise. To measure the efficiency of the search we simulate three different populations of waveforms~\cite{S5_S6_paper}. The injection parameters are chosen in such a way that they cover a wide range of signal parameter space. We utilize the injection and the slide triggers to find a balance between recovering as many simulated signals in coincidence between multiple detectors as possible while keeping the rate of false coincidences to a minimum. This process is known as the tuning of a search pipeline~\cite{S4cbctune}.

The detection statistic is a ranking device that is constructed from the SNRs of coincident triggers. The exact form of that statistic depends on the properties of the SNR distributions of injection and background triggers in a given data set. In the S4 ringdown search, no triple-coincident event was found. Nonetheless, a ranking statistic for such events was proposed~\cite{PhysRevD.80.062001}\footnote{The original definition does not have a square-root at the RHS.}:
\begin{equation}\label{eq5_oldtripcoinc}
\rho_{\mbox{\scriptsize{S4,trip}}} = \sqrt{\rho_{\mbox{\scriptsize{ifo1}}}^{2}+\rho_{\mbox{\scriptsize{ifo2}}}^{2}+\rho_{\mbox{\scriptsize{ifo3}}}^{2}}\,,
\end{equation}
where $\rho_{\mbox{\scriptsize{ifoN}}}$ is the SNR in the $N^{\text{th}}$ interferometric observatory or ``ifo'' and is equivalent to $\rho_c$ defined in Eq.~\eqref{eq5_14}.\footnote{Note that
an observer may choose to use a single-detector statistic other than $\rho_c$ for use in place of $\rho_{\mbox{\scriptsize{ifoN}}}$ in Eq.~(\ref{eq5_oldtripcoinc}).} The square-root of the entity introduced in the above equation is termed as the combined SNR.

On the other hand, the S4 ringdown search did find may double-coincident events. The following detection statistic for double-coincident events was introduced as
\begin{equation}\label{eq5_doublestat}
\rho_{\mbox{\scriptsize{S4,doub}}} = \mbox{min}\big\{\rho_{\scriptsize{\mbox{ifo1}}}+\rho_{\scriptsize{\mbox{ifo2}}}\,,\,a\rho_{\scriptsize{\mbox{ifo1}}}+b\,,\,a\rho_{\scriptsize{\mbox{ifo2}}}+b\big\}\,,
\end{equation} 
where the tunable parameters $a$ and $b$ were set to 2 and 2.2, respectively. The statistic is discussed in more detail in Ref.~\cite{PhysRevD.60.021101}. Due to the appearance of the contours of constant $\rho_{\mbox{\scriptsize{S4,doub}}}$, this statistic is also known as the ``chopped-L-stat". This may be compared with the combined SNR of a double-coincident trigger, which is just the sum of squares of the SNRs in the two detectors similar to Eq.~(\ref{eq5_oldtripcoinc}). However, the SNR distribution in real data was found to have long ``tails'', i.e, coincidences with a very loud SNR in one detector and a much lower SNR in the other, which motivated the form in Eq.~\eqref{eq5_doublestat}. When one explores the behavior of Eq.~(\ref{eq5_doublestat}), it becomes clear that a double coincidence with an H1L1 SNR pair of (10, 10) in each detector would be ranked higher than an H1L1 SNR pair of (5, 20). This is the opposite behavior of Eq.~(\ref{eq5_oldtripcoinc}). While it is true that a real gravitational wave source could have an orientation that would produce an SNR combination of (5, 20) in two non-collocated detectors, the occurrence is relatively rare compared to the number of background coincidences that could produce such a combination. Nevertheless, if one wished to search more carefully for systems with orientations leading to asymmetric SNR distributions, Eq.~(\ref{eq5_doublestat}) is insufficient and different methods of ranking would need to be employed.

\begin{figure}[!h]
\begin{center}
{\includegraphics[width=0.46\textwidth]{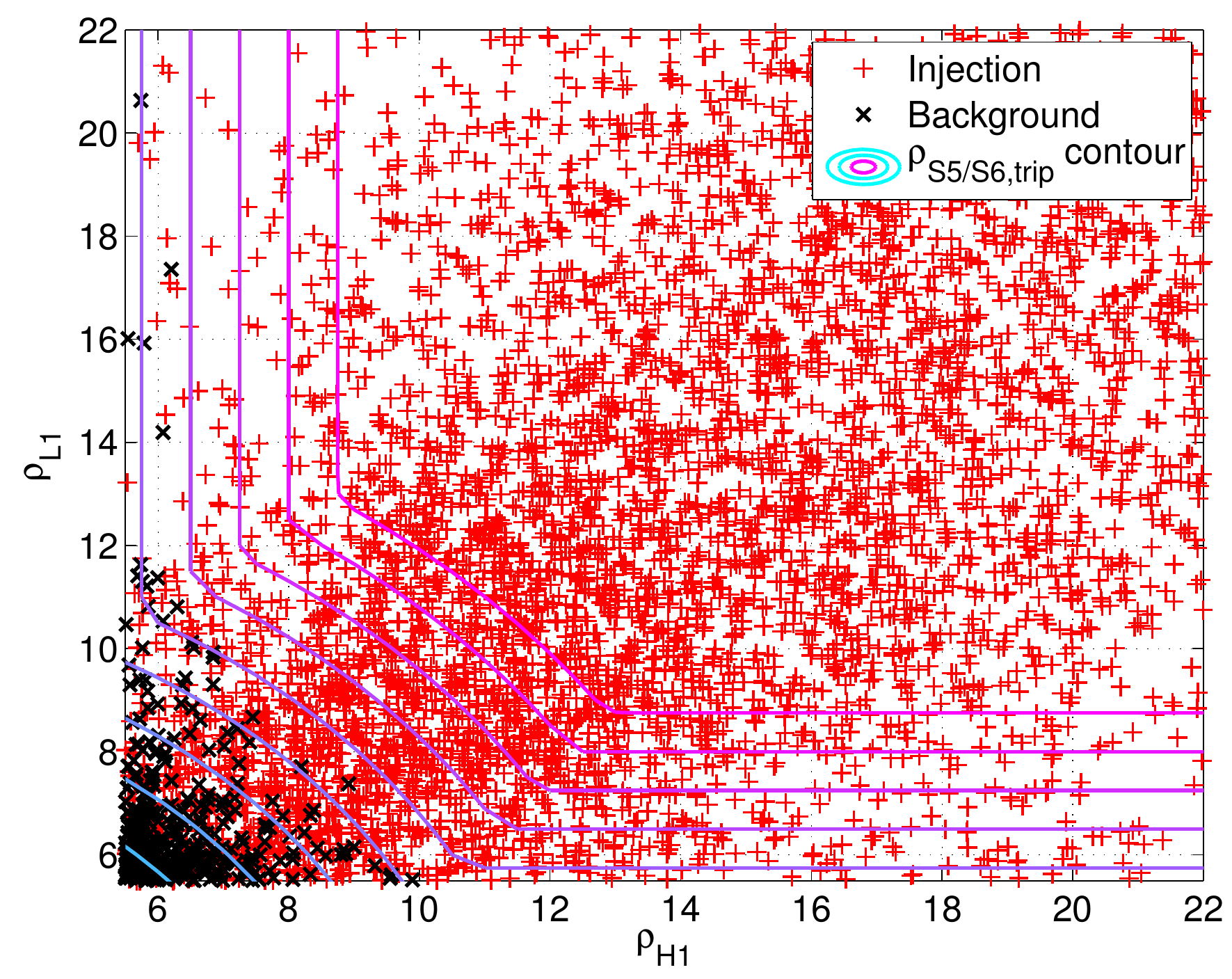}}
\caption{\label{fig:detstatcontour} The H1L1 SNR distributions for time-slide (black crosses) and injection triggers (red pluses) found coincident in H1, H2, and L1. The curves represent the contours of constant values of the new detection statistic for triple-coincident events (for $\rho_{\mbox{\tiny{H2}}}=8.0$) defined in Eq.~\eqref{eq5_newtripcoinc}.}
\end{center}
\end{figure}

For the S5/S6 ringdown search, the distribution of double-coincident triggers was found to follow the same distribution as in the S4 ringdown search where tails due to a high SNR in only one interferometer appeared. So, we continued the use of Eq.~\eqref{eq5_doublestat} to rank double-coincident triggers. However, by running the search on the significantly longer analysis time of S5, lowering the SNR threshold, and tuning the search with a better understanding of systematics, sources, and the pipeline, we managed to generate hundreds of triple-coincident background triggers. This is in contrast to the dearth of triple coincidences found in the S4 search. Furthermore, similar to what was observed for double-coincident background events, we found that triple-coincident background events also have tails due a high SNR in only one detector. This is demonstrated in Fig.~\ref{fig:detstatcontour} where we have plotted triple-coincident background events in their H1L1 SNR plane. Particularly conspicuous is the tail of background events with L1 SNR $>$ 14 and H1 SNR $<$ 7. 

These loud L1 background events are caused by noise transients in the L1 detector. The background rate for GW matched filter searches is significantly affected by the presence of noise transients (glitches). In compact binary coalescence (CBC) searches, signal-based vetoes such as the $\chi^{2}$ waveform consistency test is used to discriminate genuine events from the false ones~\cite{S4cbctune}. This test checks for how well a signal matches a template by examining its projection onto an orthogonal decomposition of the template~\cite{PhysRevD.71.062001}. It is found to work well for broadband signals 
but is not an option for monochromatic ringdown signals. Thus, we choose to again exploit the behavior of glitches and design a ``chopped-L"-like triples statistic to down-weight triggers in the tail of one detector significantly. This was done by defining a new detection statistic for triple-coincident events with contours of constant value shown in Fig.~\ref{fig:detstatcontour}:  
\begin{eqnarray}\label{eq5_newtripcoinc}
\rho_{\mbox{\scriptsize{S5/S6,trip}}} &=& \mbox{min}\Big\{\sqrt{\rho_{\mbox{\scriptsize{ifo1}}}^{2}+\rho_{\mbox{\scriptsize{ifo2}}}^{2}+\rho_{\mbox{\scriptsize{ifo3}}}^{2}}\,,\nonumber \\
&&\rho_{\mbox{\scriptsize{ifo1}}}+\rho_{\mbox{\scriptsize{ifo2}}}+c\,, \,\rho_{\mbox{\scriptsize{ifo2}}}+\rho_{\mbox{\scriptsize{ifo3}}}+c\,,\nonumber\\
&&\rho_{\mbox{\scriptsize{ifo3}}}+\rho_{\mbox{\scriptsize{ifo1}}}+c\Big\}\,,
\end{eqnarray}
where the tunable parameter $c$ was set to 0.75. This allowed us to account for glitches in the plane of two-ifo SNRs as long as the difference between the loudest and quietest trigger is significant. For example, an $\mbox{SNR}\approx 20$ glitch in H2 coincident with an $\mbox{SNR}\approx 20$ in L1 with a quiet $\mbox{SNR}\approx 5.5$ trigger in H1 would be down-weighted by a small amount. This becomes more significant as the difference increases.
Otherwise, the statistic remains the same as the sum of SNR squares defined for use in the S4 search in Eq.~(\ref{eq5_oldtripcoinc}). Again, we must raise the caveat that this statistic is not sufficient for searching for systems with orientations that produce asymmetric SNR distributions. 

In order to quantify the performance of the new statistic, we injected simulated waveforms into 
%roughly 12 months of S5 data 
data collected between 6th January 2006 and 5th January 2007 from LIGO detectors to test our ability to recover them with a ranking statistic above the highest ranked background coincidence. The waveforms used for this study included both spinning and non-spinning inspiral-merger-ringdown waveform modeled by the phenomenological method~\cite{PhysRevD.82.064016,PRL.106.241101} (PhenomB) in addition to $l=m=2$ mode ringdown waveforms from perturbed black holes. The PhenomB waveforms were distributed uniformly in total mass between 50-450~M$_\odot$ and uniformly in mass ratio between 0.1-1.0. Additionally, the spinning waveforms had uniform spins between 0.1-0.85. We designed two sets of ringdown waveforms. The first was distributed uniformly in frequency between 50-2000~Hz and in quality factor between 2.1187-20. The second was distributed uniformly in final black hole mass between 50-800~M$_\odot$ and spin between 0.1-0.99. All injections were given random sky locations and inclinations distributed uniformly over $\cos(\iota)$.

\begin{figure}[!h]
\begin{center}
{\includegraphics[width=0.45\textwidth]{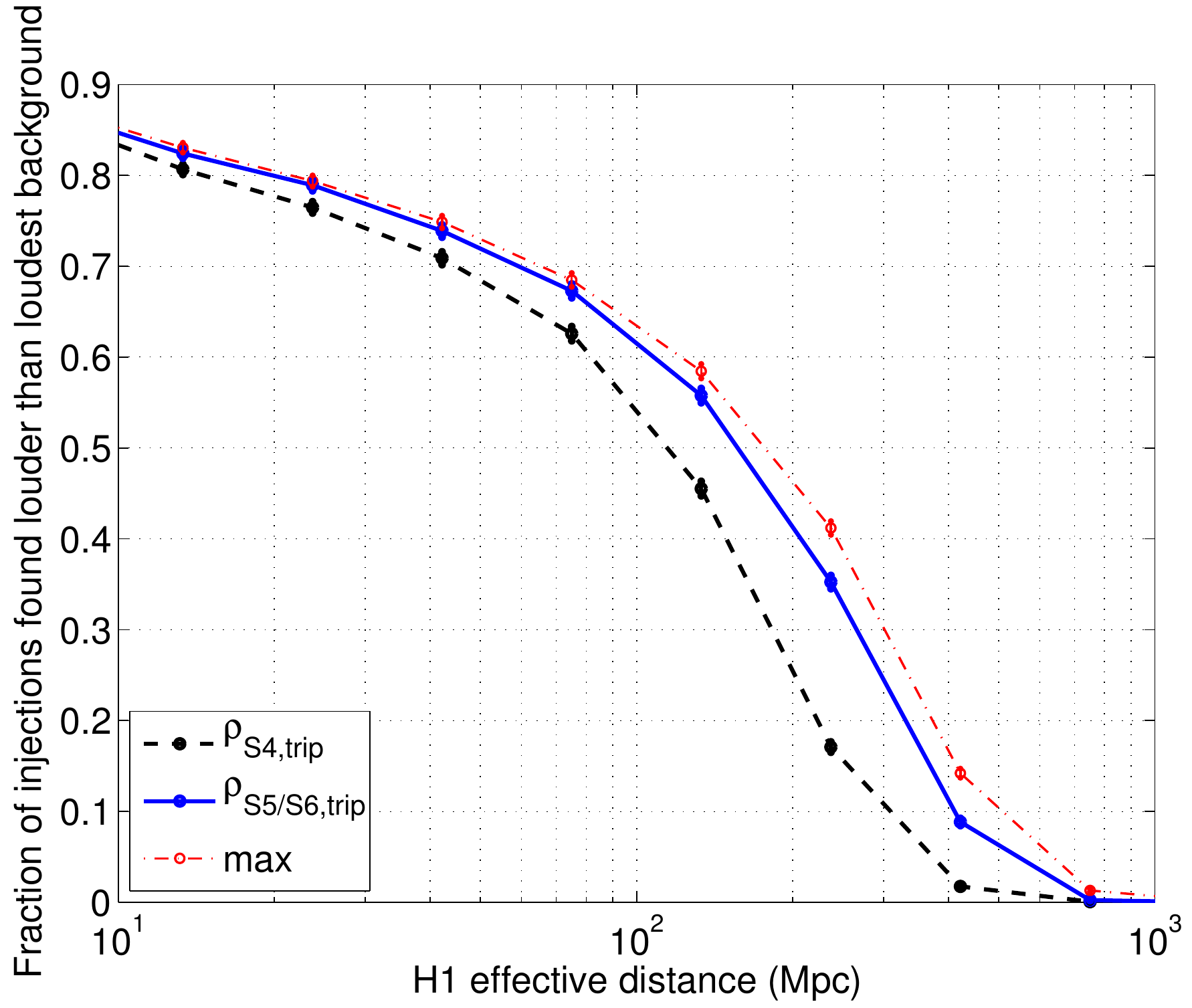}}
\caption{\label{fig:detstatefficiency} The efficiency of finding injections using the old detection statistic for triple-coincident events is compared with that of using the new detection statistic for triple-coincident events. The maximum achievable efficiency, for the injections considered here, is shown in ``max''. In this case, we assume that all found injections are louder than the loudest background.}
\end{center}
\end{figure}

The distribution of H1L1 SNR for those injections that were found in triple coincidence are shown in Fig.~\ref{fig:detstatcontour}. This includes roughly 30,000 PhenomB injections and 23,000 ringdown injections. Also plotted are 380 background coincidences. The performance of the new detection statistic for triple-coincident events is compared with the old detection statistic for triple-coincident events in Fig.~\ref{fig:detstatefficiency}. The details of the search parameters can be found in Ref.~\cite{S5_S6_paper}. Using Eq.~(\ref{eq5_oldtripcoinc}), the loudest background coincidence fell in the tail with a set of H1H2L1 SNRs of (5.74, 4.40, 20.63) and a statistic value of $\rho_{\mbox{\scriptsize{S4,trip}}}=22$. Roughly 16,900 simulated waveforms were found with a larger value of $\rho_{\mbox{\scriptsize{S4,trip}}}$. Using Eq.~(\ref{eq5_newtripcoinc}), the loudest background coincidence had a set of H1H2L1 SNRs of (6.60, 7.56, 9.44) and a statistic value of $\rho_{\mbox{\scriptsize{S5/S6,trip}}}=14$. Roughly 20,000 simulated waveforms were found with a larger value of $\rho_{\mbox{\scriptsize{S5/S6,trip}}}$. Thus a larger fraction of the injection triggers become louder than the loudest background trigger and thereby become more significant as detection candidates when using the new ranking statistic $\rho_{\mbox{\scriptsize{S5/S6,trip}}}$ as compared to using $\rho_{\mbox{\scriptsize{S4,trip}}}$. In addition, since H2 is half as sensitive as H1, we applied a cut by retaining triggers with $\rho_{\mbox{\tiny{H1}}} > \rho_{\mbox{\tiny{H2}}}$. In turn, this improves the detection efficiency appreciably. Quantitatively speaking, the detection probability of this new statistic is higher than that of the old statistic at low false-alarm probability.

\section{The coherent multi-detector statistics}\label{sec:coherent}

In this section, we develop a coherent ringdown search pipeline. We add the infrastructure needed for checking the coherence of signals from a common astrophysical source in multiple detectors to the coincidence search pipeline. As we show here, we find that the resulting search pipeline performs better than the coincidence-only ringdown pipeline. Here we detail the construction of a set of multi-detector statistics that are used in the coherent search to aid its performance in real data.

\subsection{Two-phase ringdown templates}

Traditionally, the coincidence search pipeline has not computed the quadrature phase of the ringdown signal in a detector. In fact, it has used only single-phase ringdown templates (see Eq.~\ref{eq5_13}) to filter the data. However, this limits (but does not nullify) the power of the pipeline in accessing information about the GW polarization of the signal that is available in $\varphi_0$ and using it to check for quadrature phase consistency across the detectors in a network. 

That limitation can be removed by filtering the data with both phases of the template separately~\cite{Pai:2000zt}. This straightforwardly applies to the frequency-domain inspiral signal in the stationary phase approximation~\cite{Brown:2004vh}. This is because the sine and the cosine phases of the inspiral template are exactly orthogonal. But this is not true for a generic damped sinusoid. In the following sections, we discuss how the two-phase template is used to filter the data.

%\subsubsection{Fourier transform of a generic decaying sinusoid}
%\label{sec:decayingoscillations}
\subsubsection{Fourier transform of a two-phase ringdown template}
\label{sec:FTringdown}

%\subsubsection{Fourier transform of an two-phase ringdown template}
%\label{sec:FTringdown}
The two-phase ringdown template is expressed as
\begin{equation}\label{eq_cpw}
h_{\scriptsize{\mbox{ep}}}(t) = h_{\scriptsize{\mbox{c}}}(t)-ih_{\scriptsize{\mbox{s}}}(t)\, , 
\end{equation}
where $h_{\scriptsize{\mbox{c}}}(t)$ and $h_{\scriptsize{\mbox{s}}}(t)$ are two damped sinusoid functions given by
\begin{eqnarray}\label{eq_hchs}
h_{\scriptsize{\mbox{c}}}(t) &=& e^{-\pi f_{0}t/Q}\,\cos(2\pi f_{0}t)\, ,\\
h_{\scriptsize{\mbox{s}}}(t) &=& e^{-\pi f_{0}t/Q}\,\sin(2\pi f_{0}t)\, ,
\end{eqnarray}
for $t>0$. The Fourier transform of the above functions are given by 
%The real part of $h_{\scriptsize{\mbox{ep}}}(t)$ can be found from Eq.~(\ref{eq_do}) by setting $A=1,~\alpha=\pi f_{0}/Q,~\omega=2\pi f_{0}$ and $\varphi=0$. Hence, the Fourier transform of $h_{\scriptsize{\mbox{c}}}(t)$ is found Eq.~(\ref{eq_fftx2}), and is given by
\begin{eqnarray}\label{eq_fthc}
\tilde{h}_{\scriptsize{\mbox{c}}}(f) &=& \frac{\frac{\pi f_{0}}{Q}+i2\pi f}{4\pi^{2}f_{0}^{2}-4\pi^{2}f^{2}+\frac{\pi^{2} f_{0}^{2}}{Q^{2}}+i\frac{4\pi^{2}ff_{0}}{Q}}\, ,\\
%\end{equation}
%Similarly, the imaginary part of $h_{\scriptsize{\mbox{ep}}}(t)$ is found from Eq.~(\ref{eq_do}) by setting $A=1,~\alpha=\pi f_{0}/Q,~\omega=2\pi f_{0}$ and $\varphi=\pi/2$. Hence, the Fourier transform of $h_{\scriptsize{\mbox{s}}}(t)$ is found from Eq.~(\ref{eq_fftx2}), and is given by
%\begin{equation}\label{eq_fths}
\tilde{h}_{\scriptsize{\mbox{s}}}(f) &=& \frac{2\pi f_{0}}{4\pi^{2}f_{0}^{2}-4\pi^{2}f^{2}+\frac{\pi^{2} f_{0}^{2}}{Q^{2}}+i\frac{4\pi^{2}ff_{0}}{Q}}\, ,\label{eq_fths}
\end{eqnarray}
where, the Fourier transformation of $x(t)$ is defined by
\begin{equation}\label{eq_def_FT}
\tilde{x}(f) = \int_{-\infty}^{\infty}\dt\,x(t)\,e^{-2\pi i f t}\,.
\end{equation}
By combining Eqs.~\eqref{eq_fthc} and~\eqref{eq_fths}, one gets the Fourier transform of $h_{\scriptsize{\mbox{ep}}}(t)$. Filtering the data $x(t)$ with a two-phase template $ h_{\scriptsize{\mbox{ep}}}(t; \mu_{i})$ characterized by the source parameters $\mu_{i}$ yields the {\em complex} SNR statistic 
\begin{eqnarray}\label{eq_snr_ep}
C( h_{\scriptsize{\mbox{ep}}}) &=& \frac{\langle x, h_{\scriptsize{\mbox{ep}}}\rangle}{\sqrt{\langle h_{\scriptsize{\mbox{ep}}}, h_{\scriptsize{\mbox{ep}}}\rangle}}\, ,\\
%&=& \frac{|\langle x,\{ h_{\scriptsize{\mbox{c}}}-i h_{\scriptsize{\mbox{s}}}\}\rangle|}{\sqrt{\langle h_{\scriptsize{\mbox{ep}}}, h_{\scriptsize{\mbox{ep}}}\rangle}}\, ,\\
&\equiv& \rho( h_{\scriptsize{\mbox{ep}}}) e^{i\Phi} \,,
\end{eqnarray}
where the SNR is
\begin{equation}
\rho( h_{\scriptsize{\mbox{ep}}}) = \frac{\sqrt{\langle x, h_{\scriptsize{\mbox{c}}}\rangle^{2}+\langle x,h_{\scriptsize{\mbox{s}}}\rangle^{2}}}{\sqrt{\langle h_{\scriptsize{\mbox{ep}}}, h_{\scriptsize{\mbox{ep}}}\rangle}}\,,
\end{equation}
and $\Phi$ is the quardature phase~\cite{Bose:1999pj,0264-9381-28-13-134009}.

The detailed calculation of the template normalization is as follows (see also Ref.~\cite{PhysRevD.68.102003}):
\begin{eqnarray}\label{sigma_detail}
\sigma_{\scriptsize{\mbox{ep}}}^{2} &=& \langle h_{\scriptsize{\mbox{ep}}}, h_{\scriptsize{\mbox{ep}}}\rangle\, \noQ
%&=& \langle\{h_{\scriptsize{\mbox{c}}}-i h_{\scriptsize{\mbox{s}}}\},\{h_{\scriptsize{\mbox{c}}}-ih_{\scriptsize{\mbox{s}}}\}\rangle \, \noQ
%&=& 2\int_{-\infty}^{\infty}df\,\frac{\left[\tilde{h}_{\scriptsize{\mbox{c}}}(f)-i\tilde{h}_{\scriptsize{\mbox{s}}}(f)\right] \left[\tilde{h}_{\scriptsize{\mbox{c}}}^{*}(f)+i\tilde{h}_{\scriptsize{\mbox{s}}}^{*}(f)\right]}{S(|f|)}\, \noQ 
&=& 2\int_{-\infty}^{\infty}\df\,\frac{\left|\tilde{h}_{\scriptsize{\mbox{c}}}(f)\right|^2}{S_{h}(|f|)} 
-2i \int_{-\infty}^{\infty}\df\,\frac{\tilde{h}_{\scriptsize{\mbox{s}}}(f)\tilde{h}_{\scriptsize{\mbox{c}}}^{*}(f)}{S_{h}(|f|)} \nonumber \\
&& + 2\int_{-\infty}^{\infty}\df\,\frac{\left|\tilde{h}_{\scriptsize{\mbox{s}}}(f)\right|^2}{S_{h}(|f|)} + 2i \int_{-\infty}^{\infty}\df\,\frac{\tilde{h}_{\scriptsize{\mbox{c}}}(f)\tilde{h}_{\scriptsize{\mbox{s}}}^{*}(f)}{S_{h}(|f|)}\,. \nonumber \\ \label{sigma_detail_4}
\eea
The second and the last terms in Eq.~(\ref{sigma_detail_4}) cancel each other~\cite{Talukder:2012}. So, the final expression for the template norm is given by
\begin{eqnarray}\label{sigma_detail_5}
\sigma_{\scriptsize{\mbox{ep}}}^{2}
&=& 2\int_{-\infty}^{\infty}\df\,\frac{\left|\tilde{h}_{\scriptsize{\mbox{c}}}(f)\right|^2}{S_{h}(|f|)} 
+ 2\int_{-\infty}^{\infty}\df\,\frac{\left|\tilde{h}_{\scriptsize{\mbox{s}}}(f)\right|^2}{S_{h}(|f|)}\,\noQ
&=& \langle h_{\scriptsize{\mbox{c}}},h_{\scriptsize{\mbox{c}}}\rangle + \langle h_{\scriptsize{\mbox{s}}},h_{\scriptsize{\mbox{s}}}\rangle\, .
\end{eqnarray}
Since the multi-detector statistics we study below will be constructed from combinations of the template, template norm, SNR and complex SNR in each detector, we use the detector index $I$ as a subscript on those quantities, such as in $h_{\scriptsize{\mbox{c}}I}$, $\sigma_{\scriptsize{\mbox{ep}}I}$, $\rho_{I}$, and $C_I$, to identify them as belonging to that detector. 

It is important to note that
\begin{equation}\label{diffnorm}
\langle h_{\scriptsize{\mbox{c}}},h_{\scriptsize{\mbox{c}}}\rangle \neq \langle h_{\scriptsize{\mbox{s}}},h_{\scriptsize{\mbox{s}}}\rangle\,,\;\;\;\;\;\;\;\mbox{for a range of}\;\mu_{i}\, .
\end{equation}
This is because $h_{\scriptsize{\mbox{c}}}$ and $h_{\scriptsize{\mbox{s}}}$ are not orthogonal due to the damping factor. %In the following section, we numerically justify this statement. 
%\subsubsection{Numerical analysis}

We now consider
% a bank of templates that includes both sine and cosine phases, 
the S5/S6 ringdown search template bank characterized by a range of central frequencies $f_{0}\in[50-2000]$~Hz and quality factors $Q\in[2.1187-20.0000]$. For each template, we define 
%a quantity called ``fractional difference'' as
the quantity
\begin{equation}\label{fractionalchange}
\mbox{Fractional difference} \equiv \frac{\langle h_{\scriptsize{\mbox{s}}},h_{\scriptsize{\mbox{s}}}\rangle -\langle h_{\scriptsize{\mbox{c}}},h_{\scriptsize{\mbox{c}}}\rangle}{\langle h_{\scriptsize{\mbox{c}}},h_{\scriptsize{\mbox{c}}}\rangle}\, .
\end{equation} 
Figure~\ref{fig:Frac} shows the contours of the fractional difference in the template parameter space. For higher central frequencies and lower quality factors, the fractional difference is significant. 
%CHECK: sentence below:
A small quality factor and a high frequency combine to increase the damping time-scale of the signal. This allows any amplitude difference between the two polarizations to contribute strongly to the aforementioned fractional difference.
%It is in those regions of the parameter space where the phase-difference between the two polarizations contributes / a significant fraction of the signal arises from 
Therefore, it is important to search with both template phases since their relative contributions in different detectors of a network, in the form of the quadrature phase, can be used to check how consistent the signals are in those detectors with a common ringdown source. The coherent statistic uses the matched-filter output from both template phases.
%quantifies this consistency.
% and, thus, search for ringdown signals in a network of detectors, it is necessary to model the template by incorporating both phases. 
\begin{figure}
\begin{center}
{\includegraphics[scale = 0.42]{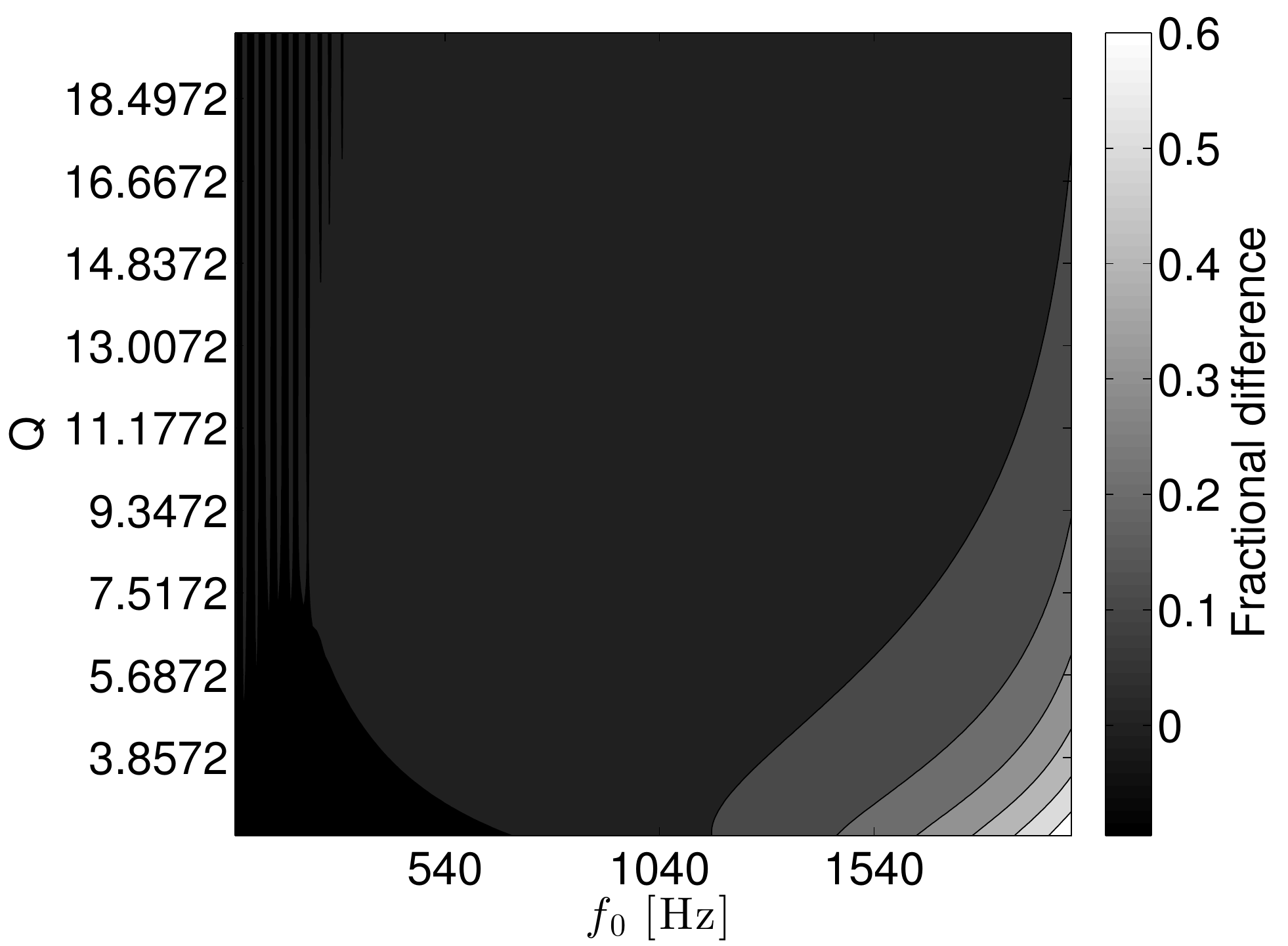}}
\caption{\label{fig:Frac} Contours of the fractional difference of sine and cosine phases of the two-phase ringdown template. The color bar represents the fractional difference as defined in Eq.~(\ref{fractionalchange}).}
\end{center}
\end{figure}

\subsection{The coherent ringdown search pipeline}
\label{app:cohpipe}
%{Working with real data}
%\label{sec:realdata}

The coherent stage is run on the coincident ringdown triggers output by the coincident multi-detector search pipeline. It has the following steps:

\noindent 1. {\em Trigbank}: This is the first step that is run in the coherent stage. It takes as input the double- and triple-coincident triggers and separates them into banks of single-detector triggers, called ``trigbank'' files.
%, groups them as double(-detector) coincidences and triple coincidences.

\noindent 2. {\em Matched filtering}: This step reads the trigbank files for each detector and filters the data from that detector using templates constructed from the parameters of each trigger listed in those files. For every trigger that crosses the chosen threshold on $\rho_I$, the signal parameters, especially, the template norm, $\rho_I$, $\Phi_I$, and the time-series of the complex SNR $C_I$ for a 125 milli-second duration centered around the trigger time, are saved. 

\noindent 3. {\em Coherent-ringdown}: This step combines the complex SNRs of a coincident trigger from each detector with appropriate weights (e.g., template-norms and, whenever applicable, time-delays and antenna factors of the different detectors) to compute the coherent detection statistic and null statistic.
 
Note that in our pipeline, the coherent detection statistic and the combined SNR computed for a double-coincident trigger are not the same, although Eq.~(\ref{eq:coinc}) may suggest so. This is because the coincident pipeline uses only the cosine template, as discussed below that equation, but the coherent search uses both templates to filter the data and compute $C_I$.

\subsection{Coherent search statistics}
\label{sec:coherentsnr}

In this section we briefly outline the basic expressions of the coherent detection statistic and the null statistics that are available for use in the ringdown search pipeline. Specifically, we focus on four categories of detector networks most relevant to the S5 coherent ringdown searches:      

\noindent
{\bf{Category I:}} {\em Two coaligned detectors with different noise PSDs.}

Let $h_{\scriptsize{\mbox{ep}}I}$ denote a two-phase template in the $I$th detector and let $\sigma_{I}$ denote its template-norm. Then,
\begin{equation}\label{eq_tmptnorm}
\sigma_{I}\equiv\sigma_{\scriptsize{\mbox{ep}}I}=\sqrt{\langle h_{\scriptsize{\mbox{ep}}I},h_{\scriptsize{\mbox{ep}}I}\rangle_I}\,,
\end{equation}
where, in keeping with Eq.~(\ref{innerprod}), we define
\be\label{innerprod_I}
\langle h_{\scriptsize{\mbox{ep}}I} , h_{\scriptsize{\mbox{ep}}I} \rangle_I = 2 \int_{-\infty}^{\infty}\df\, {\left|\tilde{h}_{\scriptsize{\mbox{ep}}I}(f)\right|^2 \over S_{h}^I(|f|)} \, , 
\ee
with $\tilde{h}_{\scriptsize{\mbox{ep}}I}(f)$ being the Fourier transform of $h_{\scriptsize{\mbox{ep}}I}(t)$ and $S_{h}^I(f)$  the noise power spectral density (PSD) of the data segment from the $I$th detector that is being filtered.\footnote{Note that the contribution to the integral in Eq.~(\ref{innerprod_I}) arises from a very narrow band, especially, for large $Q$ templates. Hence, the noise PSDs can be treated as white in computing this and similar integrals discussed later, but their values can vary with detector and template.}

The matched-filter output of the above template applied against the $I$th detector's data is:
\begin{equation}\label{eq_cdata}
C_{I}\equiv\rho_{I}e^{i\Phi_{I}}=\frac{\langle x_{I},h_{\scriptsize{\mbox{ep}}I} \rangle_I}{\sigma_I}\,,
\end{equation}
where $\Phi_I$ is the quadrature phase in the $I$th detector.
The coherent detection statistic for two coaligned detectors with different noise power spectral densities can now be defined to be
\be\label{eq_coh_2dt}
\varrho_{\underline{12}}=\frac{\left|\sigma_{1}\,C_{1}+\sigma_{2}\,C_{2} \right|}{\sqrt{(\sigma_{1})^{2}+(\sigma_{2})^{2}}}\,,
\end{equation}
where the underlined indices represent detectors that are coaligned.
On the other hand, the null statistic~\cite{0264-9381-28-13-134009} is given by
\begin{equation}\label{eq_null_2dt}
\eta_{\underline{12}} = \frac{\Big|\frac{C_{1}}{\sigma_{1}}-\frac{C_{2}}{\sigma_{2}}\Big|}{\sqrt{\left(\frac{1}{\sigma_{1}}\right)^{2}+\left(\frac{1}{\sigma_{2}}\right)^{2}}}\, .
\end{equation} 
For comparison, the detection statistic for two detectors, aligned or not, is defined as
\be\label{eq:coinc}
\rho_{12} \equiv \left( \rho_1^2 + \rho_2^2 \right)^{1/2} \,,
\ee
which is just the combined SNR, and is in line with the definition of that statistic for three detectors given in Eq.~\eqref{eq5_oldtripcoinc}. When the detectors are not aligned, the coherent statistic for a two-detector network is the same as the one given above. However in S4 and S5 the coincident search pipeline computes the above statistic (or an empirical adaptation of it, such as those discussed in Eqs.~\eqref{eq5_doublestat} and~\eqref{eq5_newtripcoinc}) with the matched-filter output of only the cosine template.

Note that the coherent statistic for the coaligned detectors obeys
\be
\label{coinccoh}
\varrho_{\underline{12}}^2 = \rho_{12}^2 - \eta_{\underline{12}}^2 \leq \rho_{12}^2 \,,
\ee
% Check the sign
which implies that the smaller the null statistic's value is for a trigger, the greater is its coherent detection statistic and the closer the latter is to the combined SNR. The expectation is that the null statistic will be closer to zero for loud enough signals and very different from zero for noise triggers of the same strength, thus allowing that statistic (and, by the above relation, the coherent detection statistic) to discern signals from noise artifacts better.

\noindent
{\bf{Category II:}} {\em Three detectors with two of them coaligned and colocated at one site and the third one located at a second site, and all with different noise PSDs.}

Let detectors $I=1,2$ be at the same site. The coherent detection statistic for this network is given by
\begin{equation}\label{eq_coh_3dt}
\varrho_{\underline{12}3} = \sqrt{(\varrho_{\underline{12}})^{2}+(\rho_{3})^{2}}\,, 
\end{equation}
where the indices 1 and 2 are underlined because they denote detectors that are coaligned and the null statistic $\eta_{\underline{12}3}$ in this case
is the same as the one defined for Category I.

\noindent
{\bf{Category III:}} {\em Three detectors at different sites with different orientations and noise PSDs.}

The antenna response functions are expressed as
\begin{equation}\label{appeq_fcfp}
\left(\begin{array}{c} F_+ \\ F_\times \end{array}\right) 
= \left(\begin{array}{cc} \cos 2\psi & \sin 2\psi \\ -\sin 2\psi & \cos 2\psi \end{array}\right) 
\left(\begin{array}{c} u \\ v \end{array}\right) \, ,
\end{equation}
with $u(\theta, \phi)$ and $v(\theta, \phi)$ being the detector orientation (and sky-position) dependent functions~\cite{0264-9381-28-13-134009}. For a network of three detectors, we introduce the following shorthand symbols for quantities involving these functions: 
\begin{eqnarray}\label{appeq_A}
{\cal A}_{12}&\equiv&\left(u_{1}v_{2}-u_{2}v_{1}\right)\, ,\\
{\cal A}_{23}&\equiv&\left(u_{2}v_{3}-u_{3}v_{2}\right)\, ,\\
{\cal A}_{31}&\equiv&\left(u_{3}v_{1}-u_{1}v_{3}\right)\, ,
\end{eqnarray} 
where, $u_{I}$ and $v_{I}$ are the $I$th detector's orientation-dependent functions. We also define ${\cal A}_{ij}=-{\cal A}_{ji}$. Then the coherent and null statistics for this network are given by
\bea\label{appeq:cohsnr}
\rho_{123} &=& \Big[\big|{\cal B}_{12}C_{1}+{\cal B}_{32}C_{3}\big|^{2}+\big|{\cal B}_{31}C_{3}+{\cal B}_{21}C_{2}\big|^{2} \noQ
&&~~+\big|{\cal B}_{23}C_{2}+{\cal B}_{13}C_{1}\big|^{2}\Big]^{\frac{1}{2}} \times \noQ
&&~\left(({\cal B}_{12})^{2}+({\cal B}_{23})^{2}+({\cal B}_{31})^{2}\right)^{-\frac{1}{2}} \, ,
\eea
and
\be\label{appeq:null}
\eta_{123} = \sqrt{\frac{\Big|\frac{C_{1}}{{\cal B}_{12}}-\frac{C_{3}}{{\cal B}_{32}}\Big|^{2}+\Big|\frac{C_{3}}{{\cal B}_{31}}-\frac{C_{2}}{{\cal B}_{21}}\Big|^{2}+\Big|\frac{C_{2}}{{\cal B}_{23}}-\frac{C_{1}}{{\cal B}_{13}}\Big|^{2}}{\left(\frac{1}{{\cal B}_{12}}\right)^{2}+\left(\frac{1}{{\cal B}_{23}}\right)^{2}+\left(\frac{1}{{\cal B}_{31}}\right)^{2}}}\, ,%\nonumber\\
\ee
where
\begin{eqnarray}\label{appeq:defB}
{\cal B}_{12}\equiv{\cal A}_{12}\sigma_{1}\sigma_{2}\, ,\\
{\cal B}_{23}\equiv{\cal A}_{23}\sigma_{2}\sigma_{3}\, ,\\
{\cal B}_{31}\equiv{\cal A}_{31}\sigma_{3}\sigma_{1}\, ,
\end{eqnarray}
denote quantities involving detector orientation, sky position, and template normalization. Like ${\cal A}_{ij}$, we also have ${\cal B}_{ij}=-{\cal B}_{ji}$.

\noindent
{\bf{Category IV:}} {\em Three coaligned detectors with different noise PSDs.}

In this category, the antenna response functions of all the detectors are identical. The corresponding coherent and null statistics are then straightforward extensions of their respective Category I forms:
{\small{
\be\label{eq_coh_3dco}
\varrho_{\underline{123}}=\frac{\left|\sigma_{1}\,C_{1}+\sigma_{2}\,C_{2} + \sigma_{3}\,C_{3}\right|}{\sqrt{(\sigma_{1})^{2}+(\sigma_{2})^{2}+(\sigma_{3})^{2}}}\,
\end{equation}
and
\begin{equation}\label{eq_null_3dc}
\eta_{\underline{123}} = \sqrt{\frac{\sigma_3^{-2}\Big|\frac{C_{1}}{\sigma_{1}}-\frac{C_{2}}{\sigma_{2}}\Big|^{2}
+\sigma_2^{-2} \Big|\frac{C_{3}}{\sigma_{3}}-\frac{C_{1}}{\sigma_{1}}\Big|^{2}
+\sigma_1^{-2} \Big|\frac{C_{2}}{\sigma_{2}}-\frac{C_{3}}{\sigma_{3}}\Big|^{2}}
{\left(\frac{1}{\sigma_{1}}\right)^{2}+\left(\frac{1}{\sigma_{2}}\right)^{2}
+\left(\frac{1}{\sigma_{3}}\right)^{2}}} \, ,%\nonumber\\
\end{equation}
}}
where, as before, the underlined indices represent detectors that are coaligned. When one of the three detectors has a much weaker sensitivity than the other two both these expressions assume the correct limit of their Category I couterparts.

%\subsubsection{Two coaligned detectors at the same site}
\subsubsection{Ringdown search in a pair of LIGO detectors}

We ran the coherent search pipeline, described in Section~\ref{app:cohpipe}, on a couple of months of S5 data from H1, H2, and L1 to study how well it does compared to the coincident search in detecting signals. Simulated ringdown signals were injected into the data of the same type as described above, for a range of quality factors and fundamental frequencies. In this paper we present results from a study where for the coherent search the statistics used took all three detectors to be coaligned. While in reality H1 and H2 are indeed coaligned, L1 is oriented slightly differently from them. That difference does allow for signals in L1 from a common source to be quite different from those in H1 for some parts of the sky. But for most of the sky they are expected to have a strong match, albeit, with a time-delay that is proportional to the projection of the LHO-LLO baseline on the direction to the source. Treating all LIGO detectors as coaligned reduces the computational cost of the search (since the pipeline no longer needs to search in the two sky-position angles but only in the time-delay along the LHO-LLO baseline) and also allows us to check how relevant a coherent analysis will be in the early Advanced Detector Era where only H1 and L1 are expected to operate. (We discuss the latter scenario in more detail in Sec.~\ref{sec:realdata}.)

The results from this study are plotted in Figs.~\ref{fig:cohvscombsnr}-\ref{fig:nullcrossnull}. They show the distribution of the values of various statistics for injection and background triggers from L1 that are coincident with H1 and H2, when the search treats L1 to be coaligned with the LHO detectors. We call this the case of the coaligned-L1 network. Therefore, when coaligned-L1 is included, all the statistics used are from network Category I for double-coincident triggers of types H1H2, H1L1, H2L1, and Category IV for triple-coincident triggers of the type H1H2L1. For this study,
% only ringdown templates were used but 
simulated injections included a variety of signal types, namely, ringdown signals and inspiral-merger-ringdown signals modeled by the phenomenological method and by the effective one-body formalism with inputs from numerical relativity (EOBNRv2)~\cite{PhysRevD.84.124052}.

Figure~\ref{fig:cohvscombsnr} bears out the inequality in Eq.~(\ref{coinccoh}): The coherent detection statistic of every trigger, be it from a signal injection or a background slide, is less than or at most equal to its combined SNR. Also, barring a few exceptions, the injection triggers mostly line up along the diagonal line where the coherent detection statistic equals the combined SNR. In other words, their null statistics are typically very small. The background triggers are more scattered. This is because their null statistics have a bigger spread in values. 

While it is nice to find these agreements between theory and experiment, the biggest drawback highlighted by this figure is that neither the combined SNR nor the coherent detection statistic is a good statistic for detecting ringdown signals, even when they are as loud as several hundred in either SNR. This is because there are very loud noise glitches that fool the ringdown search pipeline into selecting them for real signals. And unlike compact binary coalescence (CBC) signals, which have a richer time-frequency structure, the ringdown signals can not afford a chi-square test~\cite{PhysRevD.71.062001}: 
%CHECK: The following lines, ending in the footnote.
This is not due to the lack of such a structure but, rather, due to the fact that ringdown signals can be preceded by a merger and an inspiral phase, e.g., if they were from a CBC source, some parts of which may lie in the detector band. And, in practice, a ringdown filter can match the inspiral or the merger parts better (i.e., with a higher SNR) rather than the ringdown part of the signal, thereby fooling the chi-square test by giving it a larger value than what one would naively expect.\footnote{We thank Jolien Creighton for emphasizing this point to us.} 
This forces one to pursue other signal-based and detector characterization
%CHECK: cite detchar papers
tests and vetoes, especially, those that test for the consistency of the signal in multiple detectors. In this paper, we limit our attention to signal-based discriminators of that type.

One new statistic that can be derived from the coherent detection statistic of two coaligned detectors is the following cross-detector statistic:
\be
\zeta_{\underline{12}} \equiv \left( \varrho_{\underline{12}}\right)^2 - \frac{\left( (\sigma_{1})^{2}\rho_1^2 + (\sigma_{2})^{2}\rho_2^2\right)}{(\sigma_{1})^{2}+(\sigma_{2})^{2}}\,,
\ee
%CHECK: Is this related to what others have studied before?
which comprises only mixed terms in $\rho_{1}$ and $\rho_{2}\,$, hence, its name. 
%CHECK Reviewed: The following statement is not relevant to a 2D network and hence has been removed: For coaligned detectors, the same expression holds with $\zeta_{\underline{12}}$ and $\varrho_{\underline{12}}$ replacing $\zeta_{12}$ and $\varrho_{12}$, respectively.
For three coaligned detectors forming a Category IV network, the cross-detector statistic is
\be
\zeta_{\underline{123}} \equiv \left( \varrho_{\underline{123}}\right)^2 - \frac{\left( (\sigma_{1})^{2}\rho_1^2 + (\sigma_{2})^{2}\rho_2^2 + (\sigma_{3})^{2}\rho_3^2\right)}{(\sigma_{1})^{2}+(\sigma_{2})^{2} + (\sigma_{3})^{2} }\,.
\ee
Unlike the coherent detection statistic, this statistic can take positive and negative values.
%{\color{magenta}(Coh-snr for three coaligned detectors, $\varrho_{\underline{123}}$, is not defined)\color{red}[ANS: This has been corrected too.]\\}

In comparison to the coherent detection statistic, the cross-detector statistic tests how commensurate quadrature phases of the signals from a common source are in the detectors. %{\color{magenta}(only two?)}. 
A plot of this statistic versus the combined SNR is shown in Fig.~\ref{fig:combvscross} for the same experiment as in Fig.~\ref{fig:cohvscombsnr}. It is manifest that whereas none of the injections is louder than the loudest background trigger, when loudness is measured as the value of the combined SNR, a large fraction of the injection triggers have their cross-detector statistic values greater than the value that statistic has for the loudest background trigger.

The null statistic is plotted versus the coherent detection statistic in Fig.~\ref{fig:nullcross}, which qualitatively suggests that these two statistics perform only marginally better in separating the injection triggers from background triggers than the statistics used in Fig.~\ref{fig:cohvscombsnr}. Contrastingly, Fig.~\ref{fig:nullcrosszoom} suggests that the null statistic and the cross-detector statistic do quite a bit better in discriminating between those two trigger populations. This figure focuses on the weak signal region but retains all of the background triggers. 
These figures show that coincidences involving L1 have statistical properties that are very similar to the same-site, H1H2, coincidences, with one important exception. Notice how evenly the cross-detector statistic values of H1H2 background triggers are distributed around zero in Fig.~\ref{fig:nullcrosszoom}. This is not the case for the H1L1 and H2L1 background triggers. This is because for those baselines, one needs to search in the time-delay of the signals at the LHO and LLO sites, which tends to select more triggers with positive cross-detector terms since they give higher coherent detection statistic values.

The improvement found in the use of the two statistics shown in Fig.~\ref{fig:nullcrosszoom}
motivates combining their strengths into one single statistic, termed the null-cross statistic:
\be
{\mbox{Null-cross~statistic}} = \varsigma ~\left[ 1 + \left(\zeta/\varpi\right)^2 \right]^{\frac{1}{2}} - \eta^2\,,
\ee
where $\zeta$ denotes the cross-detector statistic (which, e.g., is $\zeta_{\underline{12}}$ for a coaligned two-detector network, or $\zeta_{\underline{123}}$ for a Category IV network), $\eta$ the null statistic, and $\varsigma$ and $\varpi$ are constants determined empirically by examining the receiver-operating characteristics~\cite{helstrom1960statistical} as discussed below. For the data analyzed here, $\varsigma = 7000$ and $\varpi = 600$.
Figure~(\ref{fig:nullcrossnull}) shows a plot of this statistic versus the null statistic. 

\begin{figure}[h!]
\begin{center}
{\includegraphics[width=0.52\textwidth]{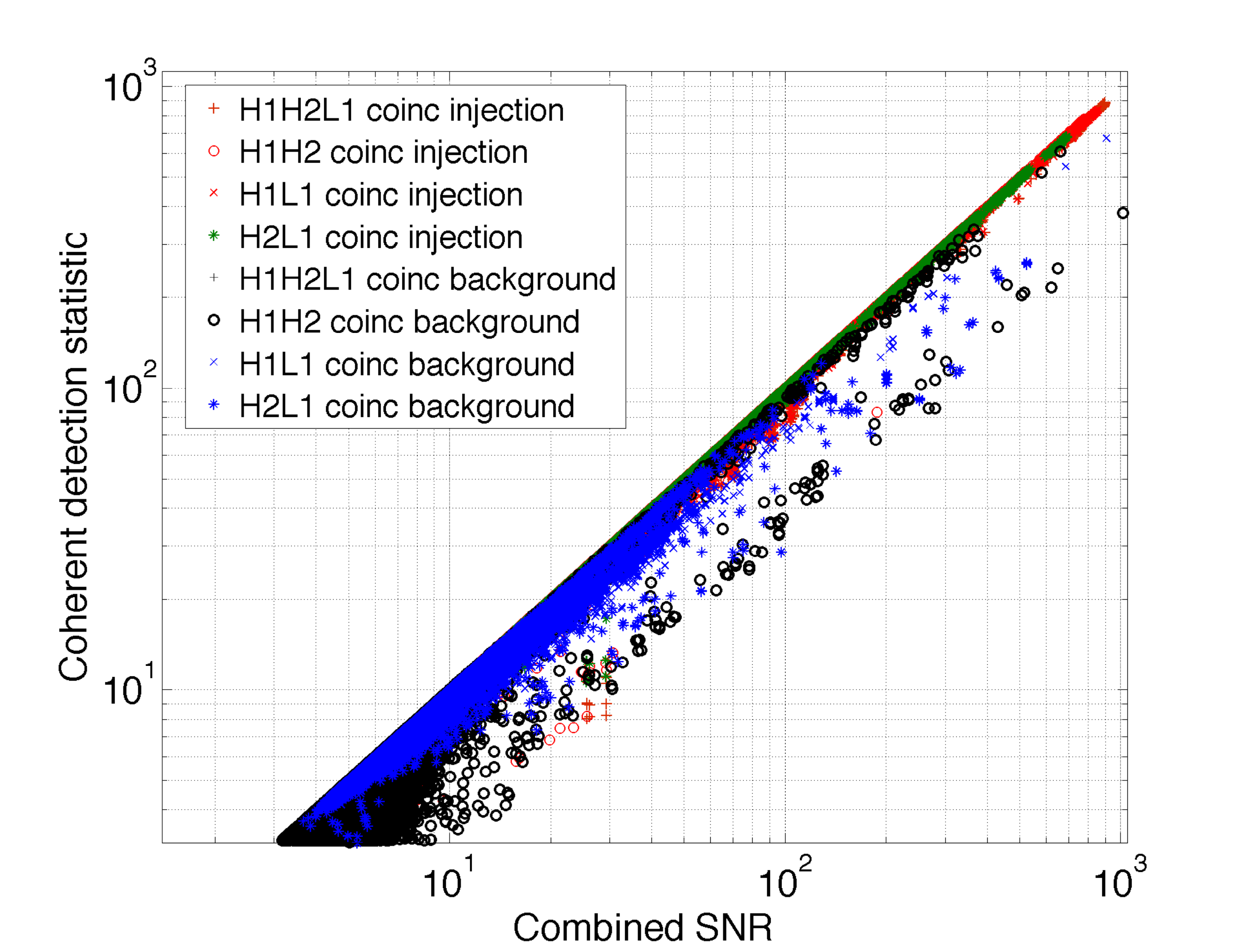}}
%{\includegraphics[scale = 0.23]{effNEWNullSqeobnrv2_lin_a_inj_h1h2l1cohslides_in_h1h2l1times_coh_snr_vs_comb_snrAll}}
\caption{\label{fig:cohvscombsnr} Double- and triple-coincident (or ``coinc'') injection and background triggers from times when all three detectors, H1, H2, and L1 had science data from months 15 and 16 of LIGO's S5 run. Notice that neither the coherent detection statistic nor the combined SNR is a good discriminator of signals since there are background triggers that are loud in both statistics.
}
\end{center}
\end{figure}

\begin{figure}[h!]
\begin{center}
%{\includegraphics[scale = 0.23]{effNEWNullSqeobnrv2_lin_a_inj_h1h2l1cohslides_in_h1h2l1times_comb_snr_vs_crossifoAll}}
{\includegraphics[width=0.52\textwidth]{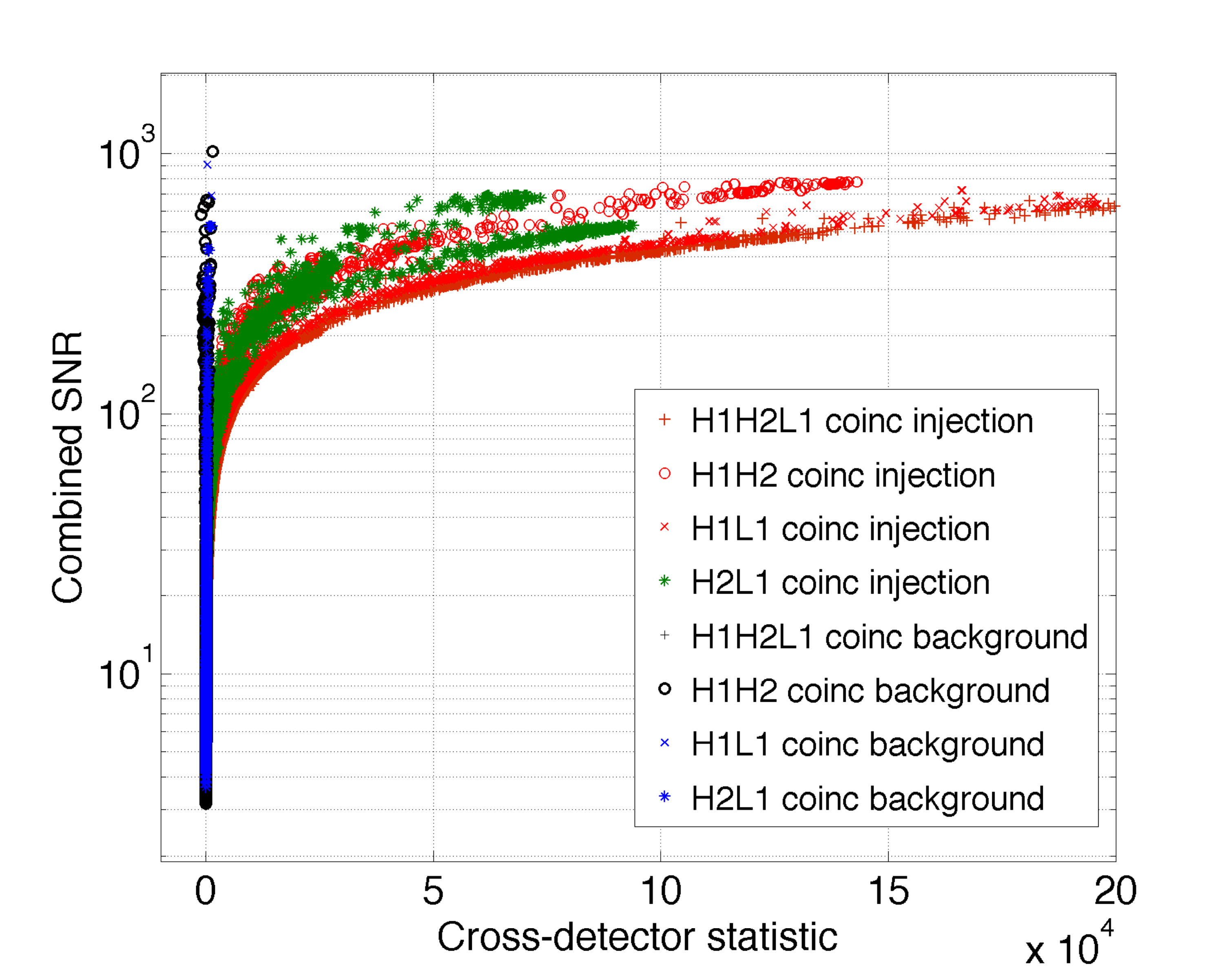}}
\caption{\label{fig:combvscross} 
The cross-detector statistic is more powerful than the combined or coherent detection statistic in discerning between signal (or injection) and noise triggers.}
%These are the same triggers as the ones shown in Fig. \ref{fig:cohvscombsnr}, and with the same legend as that figure.}
\end{center}
\end{figure}

\begin{figure}[h!]
\begin{center}
{\includegraphics[width=0.50\textwidth]{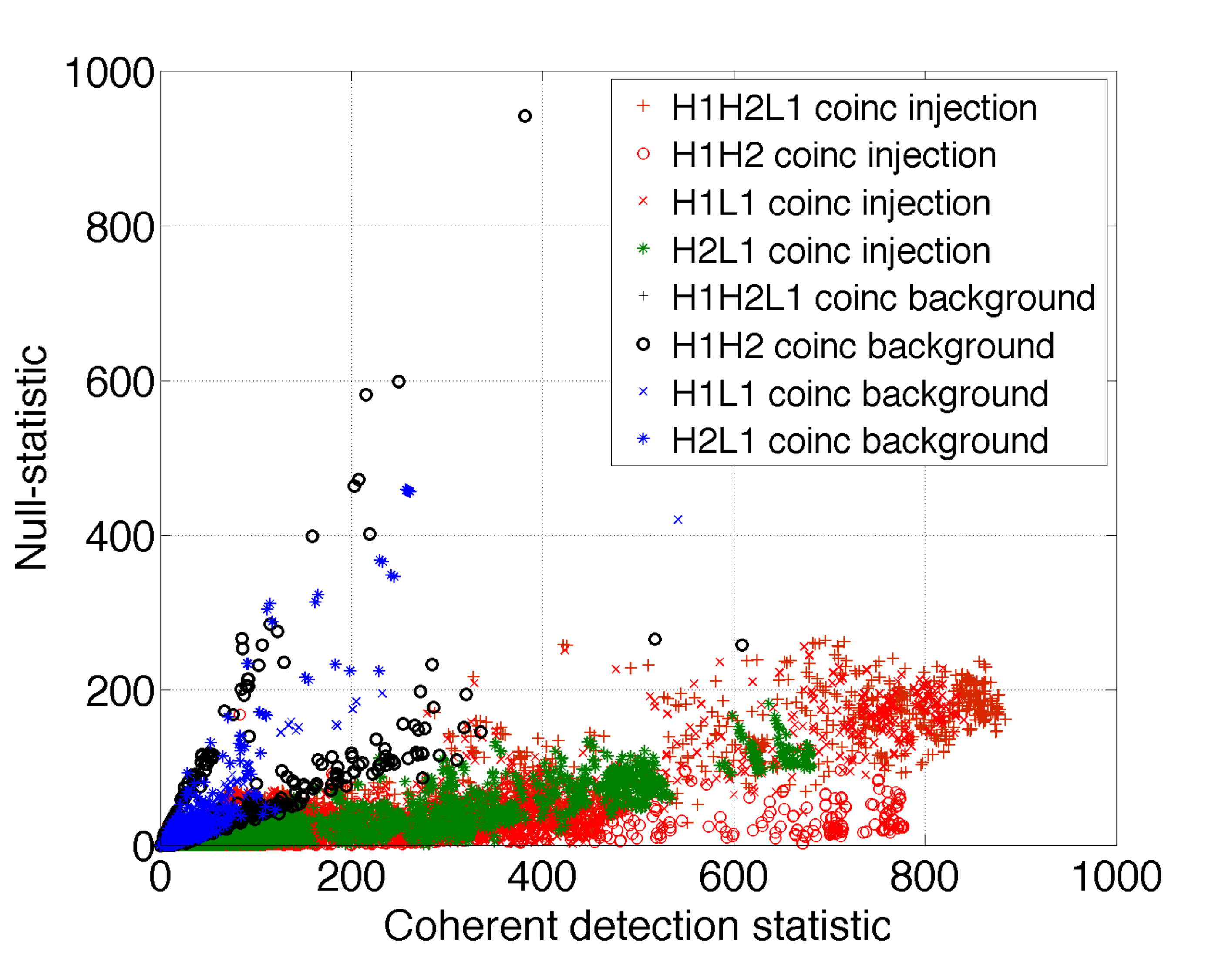}}
%{\includegraphics[scale = 0.28]{effNEWNullSqeobnrv2_lin_a_inj_h1h2l1cohslides_in_h1h2l1times_null_stat_vs_crossifoAll}}
%\caption{\label{fig:nullcross} The null and the cross-detector statistics are more successful in separating the injection triggers from the background triggers.}
\caption{\label{fig:nullcross} The null statistic and the coherent detection statistic help marginally more in separating the injection triggers from the background triggers than the statistics used in Fig.~\ref{fig:cohvscombsnr}.}
\end{center}
\end{figure}

\begin{figure}[h!]
\begin{center}
%{\includegraphics[scale = 0.28]{effNEWNullSqeobnrv2_lin_a_inj_h1h2l1cohslides_in_h1h2l1times_null_stat_vs_crossifoZoom1}}
%{\includegraphics[scale = 0.23]{effNEWNullSqeobnrv2_lin_a_inj_h1h2l1cohslides_in_h1h2l1times_null_stat_vs_crossifoZoom2}}
{\includegraphics[width=0.52\textwidth]{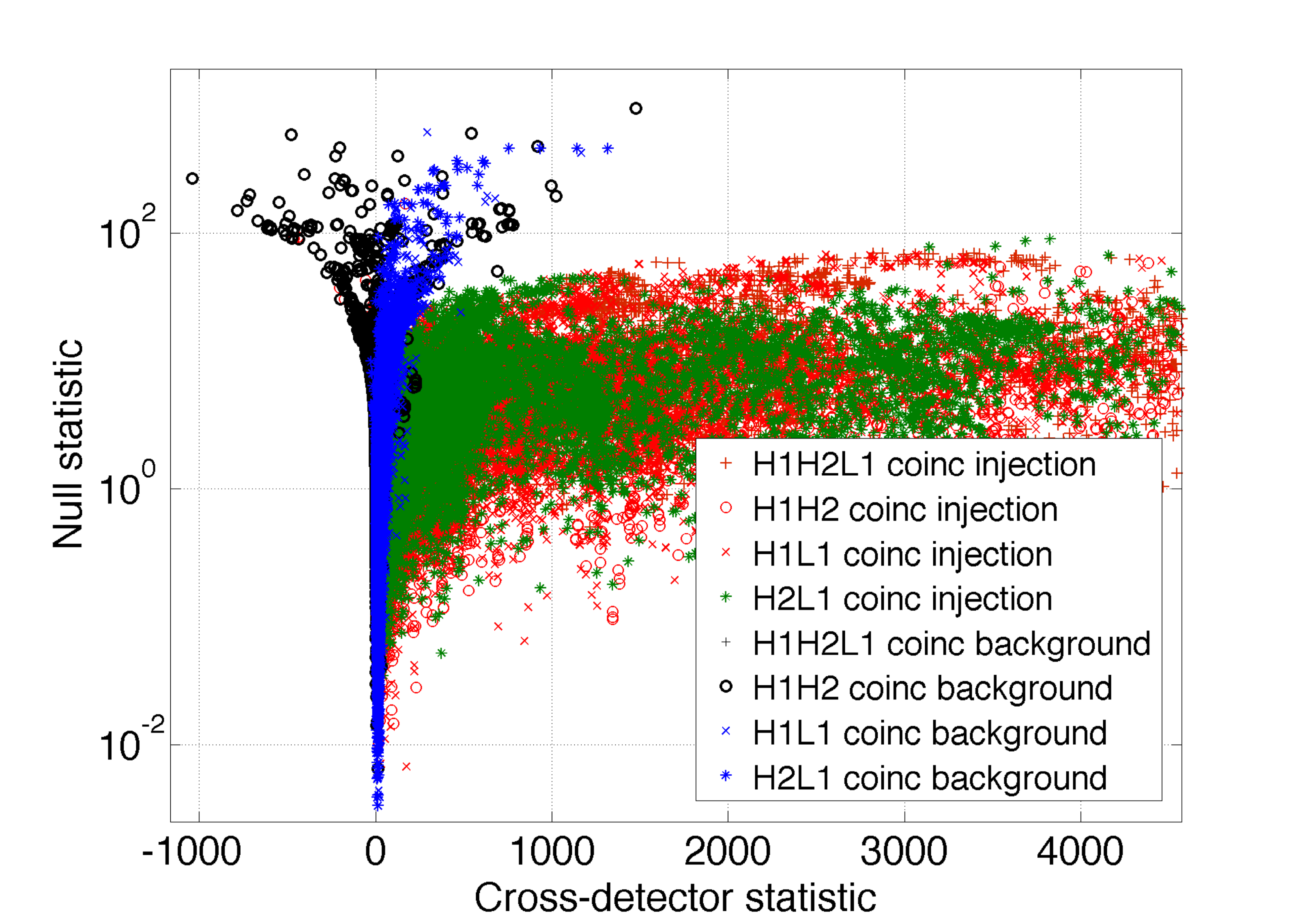}}
\caption{\label{fig:nullcrosszoom} 
A zoom of 
%the left-bottom corner of Fig. \ref{fig:nullcross} to highlight 
the region of a null statistic versus cross-detector statistic plot where the weak injection triggers reside. Notice how evenly the cross-detector statistic values of H1H2 background triggers are distributed around zero. This is not the case for the H1L1 and H2L1 background triggers. This is because for those baselines, one needs to search in the time-delay of the signals at the LHO and LLO sites, which tends to select more triggers with positive cross-detector terms since they give higher coherent detection statistic values.}
\end{center}
\end{figure}

\begin{figure}[h!]
\begin{center}
{\includegraphics[width=0.52\textwidth]{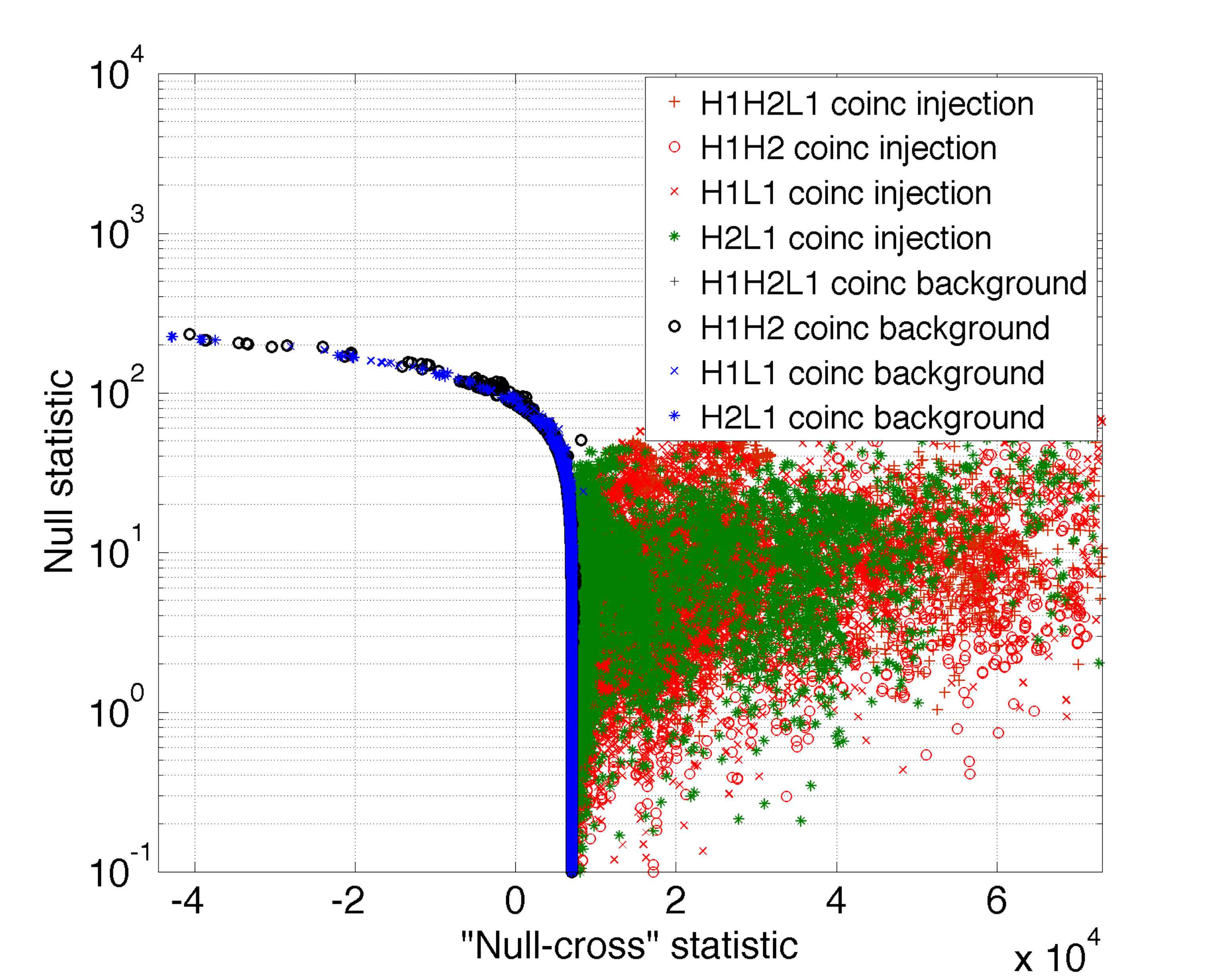}}
%{\includegraphics[scale = 0.23]{effNEWNullSqeobnrv2_lin_a_inj_h1h2l1cohslides_in_h1h2l1times_null_stat_vs_nullcrossstatZoom1}}
\caption{\label{fig:nullcrossnull} The scatter plot suggests that the ``null-cross'' statistic, which is derived by combining the null and the cross-detector statistics, might perform better in separating weak (and loud) signals from noise triggers under certain conditions. This claim is confirmed by the receiver-operating characteristic (ROC) curves in Fig.~\ref{fig:roch1l1cInh1l1_50_100}.
%CHECK: Include ROC plots
}
\end{center}
\end{figure}

\subsection{Searches in the early Advanced Detector Era}
\label{sec:realdata}

%CHECK: Should we include a plot showing the early ADE, aLIGO timelines and design curves?
It is projected that for the first few years in the advanced detector era only two LIGO detectors, one each in the Hanford (LHO) and Livingston (LLO) sites, will take science data at comparable sensitivities~\cite{ObservingScenarios}. Virgo is expected to join them with a similar sensitivity a year or so later, 
% perhaps,
%
followed by KAGRA~\cite{0264-9381-29-12-124007} and LIGO-India~\cite{LIGO_India} sometime later. It is therefore interesting to inquire if it is meaningful to pursue all-sky, all-time coherent ringdown searches in the early advanced detector era when only two detectors might be operating with the best sensitivity but at two different sites. As noted above, the similarities of H1 and L1, including their orientations, makes the case for a test that checks how consistent the signal polarizations are at the two sites. Consistency among coincident LHO and LLO triggers would increase their odds of being signals. If a source's location does not allow for this test, then it just means that it will require help from other aspects, e.g., a louder signal amplitude, to help its odds.
%CHECK: Make a sky-map showing how the match percentage, say, between plus and separately, cross polarizations vary

\subsubsection{Treating LHO and LLO as coaligned detectors}
\label{subsubcec:lholloaligned}

We now ask how the performance of a search in H1L1 data that uses a coaligned-L1 statistic, such as the null-cross statistic, compare with one that uses a regular network Category I statistic. We do so by comparing the Receiver-Operating-Characteristic (ROC) curves for the two cases in Figs.~\ref{fig:roch1l1cInh1l1full} and~\ref{fig:roch1l1cInh1l1_50_100} for H1L1 coincident triggers in H1L1 times. Each ROC curve describes how the efficiency varies as a function of its false-alarm fraction for a search with ringdown templates of simulated signals. For a given threshold value of the detection statistic, be it the combined SNR or the null-cross statistic, the false-alarm fraction is defined as the fraction of background triggers that are louder than that threshold. One expects the false-alarm fraction to monotonically decrease with increasing threshold value of the detection statistic. The efficiency at a given false-alarm fraction or, equivalently, a given threshold value of the detection statistic, 
is the fraction of signals found by the search pipeline that are louder than that threshold. One, generally, expects the efficiency to increase as one reduces the detection threshold or increases the false-alarm fraction (FAF), until one reaches a threshold above which there are no background triggers. Beyond that point, the efficiency levels off.

In Fig.~\ref{fig:roch1l1cInh1l1full} we plot the efficiency versus false-alarm fraction of a search with ringdown templates of inspiral-merger-ringdown signals, with total mass in the range 50 - 450 $M_\odot$. This figure shows that, in the FAF region where it matters the most, namely, for very low FAF values, the null-cross statistic performs better than the combined SNR. Fig.~\ref{fig:roch1l1cInh1l1_50_100} shows efficiency versus false-alarm fraction of a search with ringdown templates of ringdown signals, with final black hole mass in the range 50 - 800 $M_\odot$. The overall performance of the search with the null-cross statistic is quite a bit better than that of with the combined SNR.

\begin{figure}[h!]
\begin{center}
%{\includegraphics[scale = 0.8]{H1L1-in-H1L1-plot-FULL_DATA_CAT_4_VETO_CLUSTERED_CBC_RESULTS-852393970-4838400_ROC.png}}
{\includegraphics[scale = 0.45]{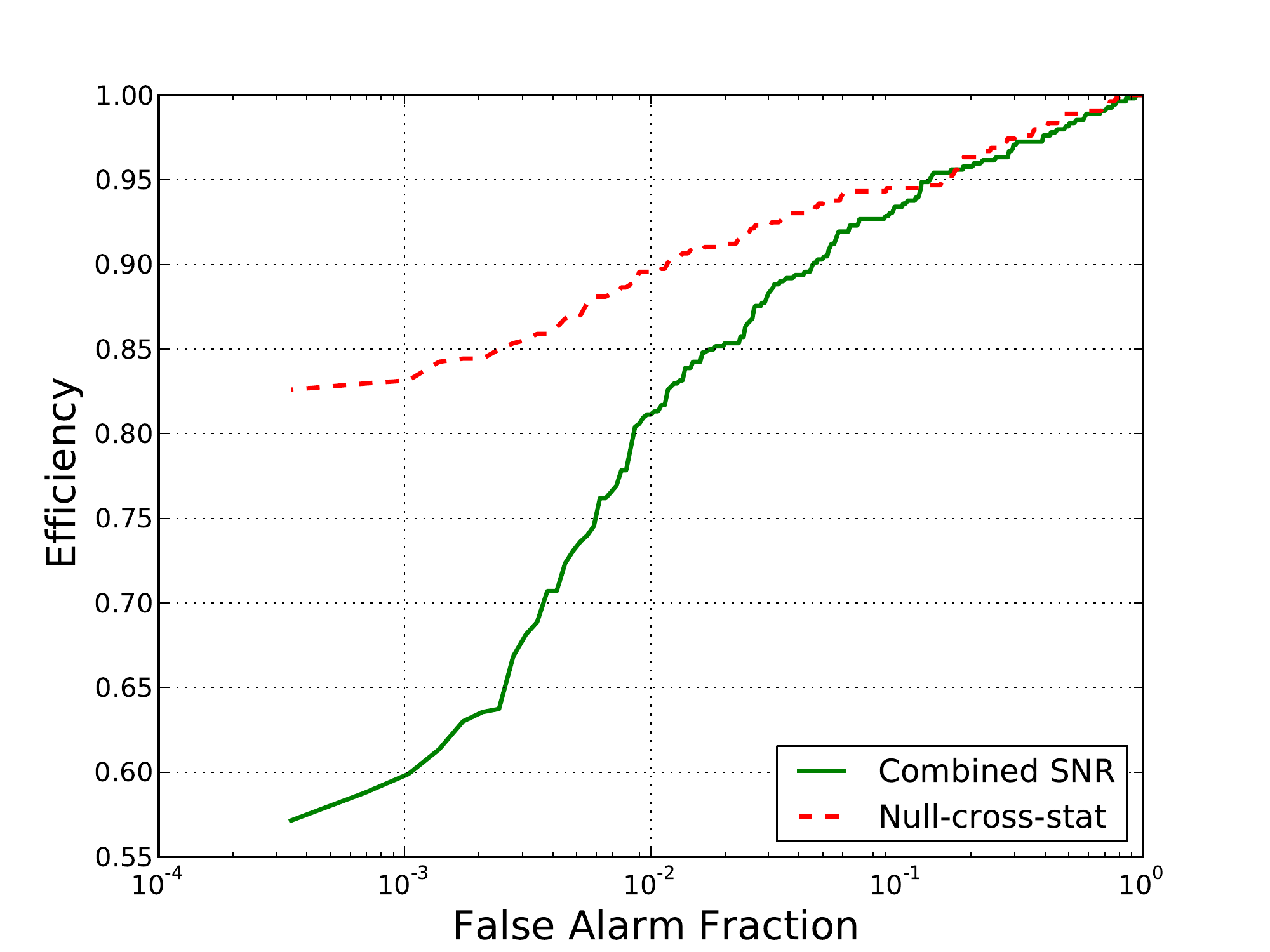}}
\caption{\label{fig:roch1l1cInh1l1full} A comparison of the ROC curves of a search with combined SNR (shown as a solid green curve) and the null-cross statistic (shown as a red-dashed curve) as detection statistic. This study used simulated inspiral-merger-ringdown signals in H1L1 data from two months of S5. Those simulations correctly accounted for the different orientations of L1 and H1 but the null-cross statistic used here was the one for the Category I network for coaligned detectors. 
}
\end{center}
\end{figure}

\begin{figure}[h!]
\begin{center}
%{\includegraphics[scale = 0.8]{H2L1-in-H2L1-plot-FULL_DATA_CAT_4_VETO_CLUSTERED_CBC_RESULTS-852393970-4838400_ROC.png}}
{\includegraphics[scale = 0.45]{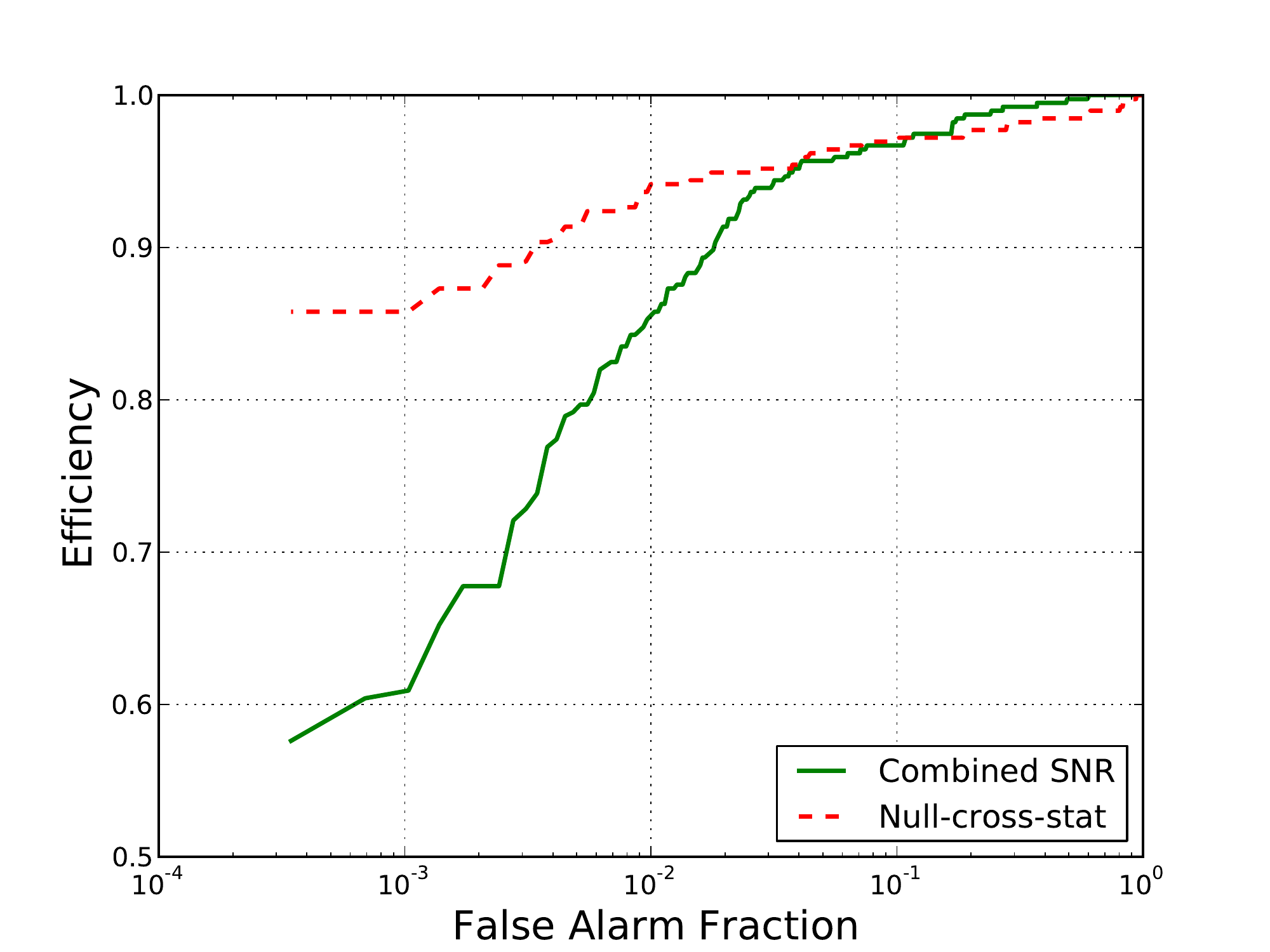}}
\caption{\label{fig:roch1l1cInh1l1_50_100} Same as Fig.~\ref{fig:roch1l1cInh1l1full} but for simulated ringdown signals.}
\end{center}
\end{figure}

\section{Discussion}
\label{sec:discussion}

The detection of intermediate mass black holes will provide support for the hypothesis that they seed the formation of super-massive black holes, many of which are known to exist in galactic nuclei. Gravitational wave detectors like LIGO and Virgo, which are currently being upgraded to achieve higher sensitivities than their first generation versions, will have the ability to detect ringdown signals from IMBHs for a wide range of masses, e.g., after they merge with other compact objects such as neutron stars, stellar mass black holes or other IMBHs. 

As we explored here, however, one of the toughest challenges in detecting these signals is posed by the detector itself, in the form its response to non-stationary instrumental and environmental disturbances. For a substantial fraction of those disturbances, the imprint of the detector's response in its data has characteristics of a damped sinusoid. The ringdown signals are very difficult to discriminate against such noise glitches. One particular source of such artifacts is the mechanical relaxation of the test-mass suspension framework in a detector. 

While efforts are underway to mitigate the occurrence of these noise glitches, in some cases, by improving the detector hardware itself, a less perfect alternative is to identify them in the analysis of the detector data. In this paper, we explored the alternative method and showed that signal consistency tests exist when employing multiple detectors that are more effective in distinguishing black hole ringdown signals in the presence of noise glitches than tests that are optimal in stationary data.
Specifically, we extended LIGO's S4 analysis~\cite{PhysRevD.80.062001} by showing that requiring conformity of the signal strengths in different detector with what we expect of real sources improved the performance of our ringdown search pipeline. Additionally, requiring the consistency of the signal phases in those detectors by effecting multi-baseline tests in the form of the null-statistic and null-cross statistic, which is derived from the coherent detection statistic, improves the performance of that pipeline in certain sections of the signal parameter space. This was demonstrated in the case where the detectors are very nearly coaligned but are located at widely separated sites. These statistics are found to be useful parameters in multivariate statistical classifier (MVSC) and be used in future searches~\cite{MVSC_paper}. Future efforts should target implementing these methods and making them effective for non-coaligned detectors located in multiple sites.

\acknowledgments 

We would like to thank Jolien Creighton for his careful reading of the manuscript and for making several important comments and suggestions for improving it. 
We also thank Matthew Benacquista, Collin Capano, Neil Cornish, Thilina Dayanga, Raymond Frey, Gabriela Gonzalez, Kari Hodge, Greg Mendell, Fred Raab, and Alan Weinstein for extensive discussions. This work was supported in part by NSF grants PHY-0855679, PHY-0855686, PHY-0970074, PHY-1205952, and PHY-1206108.

\bibliography{cohringdown}

\end{document}